\documentclass[twocolumn]{rbef}
\numeracao{xxxx}
\volume{xx}
\numero{xx}
\ano{2017xx}
\doi{http://dx.doi.org/xxxx}
\tipodeartigo{Artigos Gerais}
	\author[*]{J. A. S. Lima}
	\author[**]{R. C. Santos}
	\affil[*]{Universidade de São Paulo, Departamento de Astronomia (IAG-USP) \thanks{\href{emailto:jas.lima@iag.usp.br}{jas.lima@iag.usp.br}}
		} 
		\affil[**]{Universidade Federal de São Paulo, Campus Diadema, Departamento de Física}

\titulo{100 Anos da Cosmologia Relativística (1917-2017). Parte I: Das Origens à Descoberta da Expansão Universal (1929)}

\begin{document}

\begin{primeirapagina}

\begin{abstract}
Estamos vivenciando um período de extrema efervescência intelectual na área da cosmologia. Um volume enorme de dados observacionais em quantidade e qualidade sem precedência e um arcabouço teórico mais consistente, impeliram a cosmologia para uma era da precisão, transformando a disciplina em área de ponta da ciência contemporânea. Observações com Supernovas do tipo Ia (SNe Ia) mostraram que o Universo está expandindo aceleradamente, um fato sem explicação no modelo desacelerado tradicional. Identificar a causa dessa aceleração é hoje o problema mais fundamental da área. Tal como ocorreu no renascimento científico, a solução norteará os rumos da disciplina no futuro próximo e as possíveis respostas (se energia escura, alguma extensão da relatividade geral ou um mecanismo ainda desconhecido) deverão também alavancar o desenvolvimento da física. Nesse contexto, sem abdicar de uma abordagem pedagógica, apresentamos uma visão geral tanto dos principais resultados teóricos quanto das descobertas observacionais mais significativas da cosmologia nos últimos 100 anos. A saga da cosmologia moderna será apresentada numa trilogia. Neste artigo (\textit{Parte I}), com base nos artigos de Einstein, de Sitter, Friedmann, Lemaître e Hubble, descreveremos  o  período decorrido entre as origens da cosmologia e a descoberta da expansão Universal (1929). Na \textit{Parte II} veremos o período de 1930-1997, finalizando com o velho modelo padrão desacelerado. A \textit{Parte III} será inteiramente dedicada ao modelo acelerado do Universo,  o  paradigma cósmico do século XXI.      
\end{abstract}

\begin{abstract}
\begin{center}
\textbf{\LARGE 100 Years of Relativistic Cosmology (1917-2017). Part I: From Origins to the Discovery of Universal Expansion (1929)}
\end{center}
\vspace{0.5cm}
We are experiencing a period of extreme intellectual effervescence in the area of cosmology. A huge volume of observational data in unprecedented quantity and quality and a more consistent theoretical framework propelled cosmology to an era of precision, turning the discipline into a cutting-edge area of contemporary science. Observations with type Ia Supernovae (SNe Ia), showed that the expanding Universe is accelerating, an unexplained fact in the traditional decelerated model. Identifying the cause of this acceleration is the most fundamental problem in the area. As in the scientific renaissance, the solution will guide the course of the discipline in the near future and the possible answers (whether dark energy, some extension of general relativity or a still unknown mechanism) should also leverage the development of physics. In this context, without giving up a pedagogical approach, we present an overview of both the main theoretical results and the most significant observational discoveries of cosmology in the last 100 years. The saga of cosmology will be presented in a trilogy. In this article (\textit{Part I}), based on the articles by Einstein, de Sitter, Friedmann, Lemaître and Hubble, we will describe the period between the origins of cosmology and the discovery of Universal expansion (1929). In \textit{Part II} we will see the period from 1930 to 1997, closing with the old standard decelerated model. The Part III will be entirely devoted  to the accelerated model of the universe, the cosmic paradigm of the XXI  century. 
\end{abstract}
\end{primeirapagina}


\section{Introdução}
Cosmologia é a disciplina que investiga a forma\c{c}\~ao e evolu\c{c}\~ao da estrutura de grande escala do Universo (gal\'axias,  aglomerados e superaglomerados de gal\'axias). Seu objetivo é estabelecer um modelo cosmológico que prediga e explique os resultados das observa\c{c}\~oes astronômicas. A busca maior é a constru\c{c}\~ao de um cen\'ario que permita reconstituir o passado do Universo, entender o presente e conjecturar sobre o futuro do \textit{Cosmos}.

O estudo da cosmologia é atualmente baseado na melhor teoria de gravitação existente,  a relatividade geral de Albert Einstein. Seu progresso depende por um lado, dos desenvolvimentos teóricos nas áreas de física e astronomia e, por outro, de observações extremamente sofisticadas que podem atualmente ser consideradas como verdadeiros experimentos astronômicos. 

Na \textbf{Figura 1}, apresentamos uma cronologia dos principais  avanços teóricos e descobertas observacionais da cosmologia nos últimos 100 anos. Após um início esplendoroso onde destacamos \cite{TF71,JB86}: (i) o primeiro modelo cosmológico relativístico obtido por Einstein (1917) \cite{E17}, (ii) o modelo expansionista de Friedmann \cite{F22}, e (iii) a descoberta da expansão do Universo por Hubble \cite{H1}, a área progrediu lentamente até meados dos anos 60. Podemos dizer que  o intervalo entre 1929 e 1965 foi um período de ciência normal no sentido de Khun \cite{K}, uma espécie de ``idade média" \, tardia da cosmologia moderna; culminando numa sequência de descobertas fundamentais. 

Em 1965, Arno Penzias e Robert Wilson \cite{PW65} detectaram a radiação cósmica de fundo\footnote{A temperatura da radiação é cerca de 3K e seu pico encontra-se na banda de microondas. Na literatura inglesa é conhecida pelo acrônimo CMB (cosmic microwave background).} (RCF) e foram premiados com o Nobel de 1978. Foi um verdadeiro salto em direção à cosmologia física. 

As condições gerais reinantes nos estágios iniciais da evolução cósmica (na chamada era da radiação) foram enfim compreendidas quase completamente. Em particular, foi possível determinar teoricamente, a abundância dos núcleos atômicos leves (D, $He^{3}$, $He^{4}$ e $Li^{7}$), formados por fusão termonuclear durante esta fase; um processo que ficaria conhecido como nucleossíntese primordial\cite{Nucleo}.  

No início dos anos 70, veio a confirmação da existência de matéria escura, através das curvas de rotação de galáxias analisadas por Keneth Freeman \cite{KF70},  Vera Rubin \cite{VR} e outros. A matéria escura não brilha, sendo identificada apenas por seus efeitos gravitacionais.  Como veremos na \textit{Parte II},   a principal candidata para essa componente é uma substância inerte e não-relativística, usualmente denominada matéria escura fria ("cold dark matter - CDM). Posteriormente, a matéria escura fria,   supostamente de origem primordial,  tornou-se um ingrediente chave na compreensão do processo de formação de estruturas no Universo \cite{FE}.  

\begin{figure}
\centering
\includegraphics[width=3.3truein,height=2.9truein]{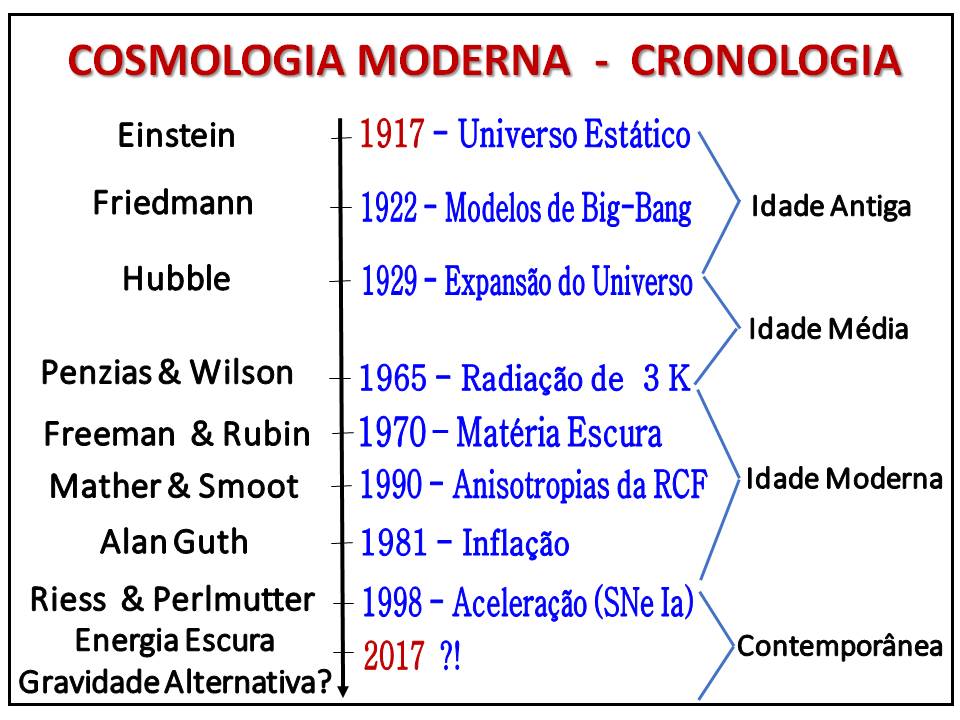}
\label{Fig0}
\caption{Linha do tempo mostrando as principais avanços teóricos e observacionais da Cosmologia Moderna \cite{BJP}.} 
\end{figure}

Em 1990, as medidas do espectro planckiano e as flutuações da RCF, obtidas com dados do satélite  COBE (\textit{cosmic background explorer}), premiaram os líderes do experimento, John Mather e George Smoot \cite{COBE} com o Nobel da Física de 2006.  O levantamento do espectro planckiano completo foi o passo definitivo  para confirmar a história térmica do Universo. Esse primeiro levantamento das flutuações intrínsicas da RCF (o termo de quadrupolo),  foram também essenciais para mostrar que  a idéia de inflação, proposta no início dos anos 80 por Alan Guth \cite{Guth}, se tornaria um elemento chave para entender os estágios mais primitivos do Universo. 

Tais descobertas deram sustentação para um cenário cosmológico expansionista, desacelerado e preenchido por matéria não-relativística (escura + bariônica); o modelo cosmológico padrão até 1997. 

Em 1998, dois grupos distintos de pesquisadores  utilizando SNe Ia como vela padrão,  mostraram que o Universo está expandindo aceleradamente \cite{R1,P1}. Esta descoberta  tornou a busca de um novo paradigma cosmológico uma tarefa inadiável \cite{BJP,Waga05}. Seus líderes, Saul Perlmutter (\textit{Supernova Cosmology Project})  e  Bryan Schmidt \& Adam Riess (\textit{High-z Supernova Search Team}),  compartilharam o prêmio nobel de 2011. 

Em 2001,  foi lançado o satélite WMAP (``Wilkinson Microwave Anisotropy Probe"), o sucessor do COBE.  Os dados liberados em 2003, mostraram o espectro completo das flutuações de temperatura (espectro angular de potências). A combinação de Supernovas com o WMAP (e depois com os dados do satélite Planck\cite{Planck}), além de uma bateria de testes, incluindo: (i) as oscilações acústicas dos bárions, (ii) a idade do Universo, (iii) dados de raios-X provenientes dos aglomerados de galáxias, e (iv) lentes gravitacionais representou um enorme avanço. Em particular, determinou-se um novo ``modelo cosmológico padrão", comumente referido como modelo de concordância cósmica (MCC) ou $\Lambda$CDM, onde $\Lambda$  é a constante cosmológica ou densidade de energia do vácuo, introduzida por Einstein em 1917 (ver seção 5). 

Tais contribuições sugerem uma subdivisão na linha do tempo da cosmologia moderna em 4 per\'{\i}odos distintos, marcados por avanços e/ou profundas mudanças na visão do \textit{Cosmos}: \textit{ Idade Antiga},  \textit{Média}, \textit{Moderna}  e \textit{ Contemporânea ou do Século XXI} (ver \textbf{Figura 1}). 

O MCC ou $\Lambda$CDM,  está de acordo com as observações existentes, mas permanece misteriosa a verdadeira causa da aceleração recente do Universo, pois a densidade de energia do vácuo ($\Lambda$) não é medida diretamente. Portanto, a existência de duas componentes desconhecidas (matéria e energia escuras)  é  uma circunstância muito desconfortável, tanto do ponto de vista teórico quanto observacional. Portanto, apesar do sucesso,  é também um momento de dúvidas e conflitos numa disciplina  centenária. 

Por outro lado, as dificuldades apontadas devem ser encaradas como uma grande oportunidade para os espíritos mais criativos. Tal como no renascimento científico, sua elucidação norteará os rumos da disciplina no futuro próximo e as respostas farão avançar o desenvolvimento da física \cite{Lima08}. 

Nesse contexto, sem esquecer aspectos pedagógicos, acreditamos ser instigante para graduandos e professores do ensino médio ter um panorama da área, seus principais resultados teóricos e as descobertas observacionais mais significativas dos últimos 100 anos\footnote{As novas idéias e descobertas, tanto quanto possível, serão apresentadas na sequência historicamente estabelecida.}. 

Tendo em vista a quantidade de eventos marcantes, dividimos o material disponível em 3 artigos subsequentes (trilogia). Na \textit{Parte I}, cobrimos desde as origens até a descoberta da expansão universal (1929). Na \textit{Parte II} veremos o período de 1930-1997, fechando com o modelo padrão desacelerado. A \textit{Parte III} será inteiramente dedicada ao modelo acelerado do Universo, o  paradigma cósmico do século XXI.    

O presente artigo \textit{(Parte I)} está organizado como segue. Na seção 2, identificamos no legado cosmológico newtoniano  o que ainda era relevante no século XX. Na seção 3, apresentamos as sementes teóricas e observacionais da cosmologia moderna. Na seção 4, discutimos as idéias básicas da relatividade geral; o alicerce  da cosmologia moderna. A seção 5 é dedicada ao modelo estático de Einstein e ao significado físico da constante cosmológica, $\Lambda$. Nas seções 6 e 7, revisamos a previsão teórica do ``big-bang" \, e a descoberta da expansão universal. Na seção 8, contextualizamos as contribuições não imediatamente repercutidas de Lemaître e, finalmente, na seção 9, apresentamos as principais conclusões dessa \textit{Parte I}. 

\section{O Legado Cosmológico Newtoniano}
\textit{``I have proposed the force of gravity but I have not yet assigned a cause to gravity" (Isaac Newton)\footnote{``Eu propus a força da gravidade, mas eu ainda não atribuí uma causa à gravidade".}}
\vskip 0.2cm

A física e astronomia estudadas hoje foram profundamente impactadas pelas leis de movimento e gravitação universal de Isaac Newton (1643-1727).  Sabemos que a possibilidade de fazer cosmologia muito  al\'em do heliocentrismo copernicano, das leis de Johannes Kepler (1571-1630) e das observações de Galileo Galilei (1564-1642), foi um legado da gravitação newtoniana.

\begin{figure}
\centering
 \includegraphics[width=2.9truein,height=2.7truein]{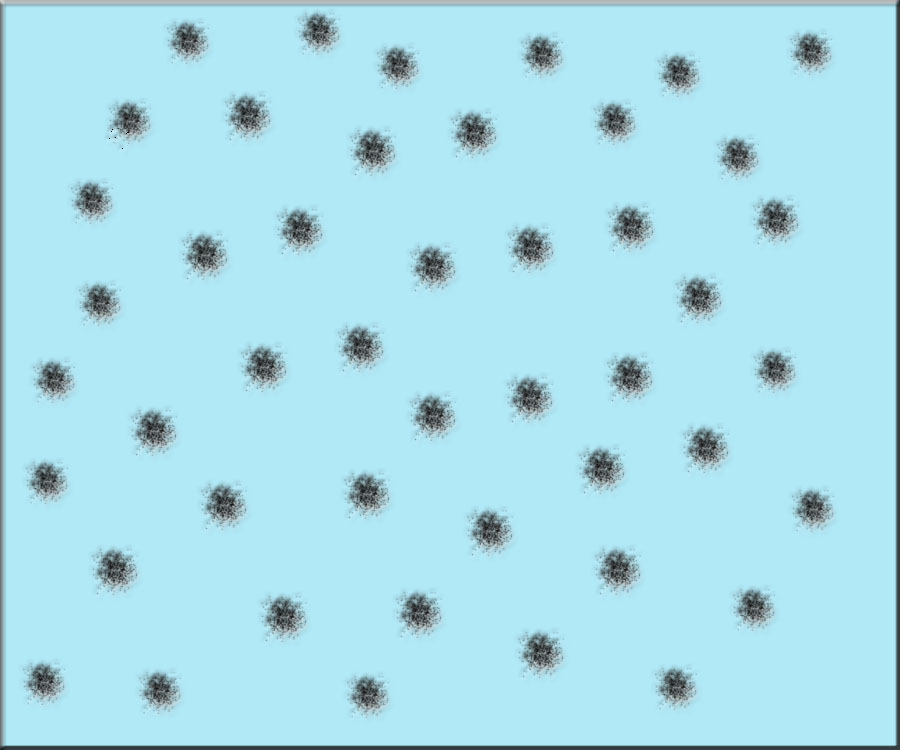}
\caption{Cosmologia pós-renascentista. Na visão de Newton, o Universo seria estático, infinito e euclidiano, com a matéria localmente agregada pela atração gravitacional. Uma visão que predominou até o século XX (ver texto).}\label{UnivNewton}
\end{figure}

Combinando suas leis da mecânica e gravitação universal,  Newton estabeleceu as bases para o desenvolvimento futuro da ciência (em várias direções e por mais de 2 séculos),  alterando profundamente n\~ao apenas a no\c{c}\~ao vigente de Universo, mas tamb\'em a própria arte de inventar ou criar teorias científicas. 

Newton deduziu todas as leis de Kepler do movimento planetário e também previu novos e importantes resultados.  Para exemplificar, consideremos a primeira  lei de Kepler ou lei das órbitas: os planetas giram em torno do Sol descrevendo órbitas elípticas (1609). Sem fazer observações, Newton demonstrou que a curva traçada pelos corpos celestes sob a\c{c}\~ao da gravidade seria em geral representada por cônicas (c\'{\i}rculo, elipse, par\'abola e hip\'erbole), explicando  simultaneamente o movimento dos planetas, satélites, cometas e demais aster\'oides \cite{KLivro}. Um exemplo antigo e marcante de como a teoria permite ver bem além do empirismo em termos de fundamentação  e compreensão lógica.
 
A segunda lei de Newton e a gravitação universal, garantiram que os ``graves" não precisavam  cair em direção ao centro, na eterna busca aristotélica de um ``lugar natural". Na verdade, planetas e sat\'elites sustentam suas órbitas em torno de centros atratores devido as forças centrípetas de origem puramente gravitacional. Uma explicação matemática rigorosa para a surpresa que teve Galileu e a elite intelectual da época, quando o pai da ciência moderna apontou sua luneta e observou os 4 maiores satélites de Júpiter. 

Newton também ampliou nossa visão do Universo. Diferente de Copérnico e do \textit{Cosmos} grego\footnote{Desde Aristóteles o Universo era considerado finito e limitado pela ``esfera das estrelas", uma  visão também compartilhada por Copérnico e Kepler \cite{EH95}.}, o Universo observado para Newton seria espacialmente infinito e dotado a priori de uma geometria euclidiana 

Numa de suas famosas cartas \cite{4NBL} respondendo algumas indagações do escolástico e reverendo Richard Bentley (1662-1742), Newton sintetizou sua vis\~ao cosmol\'ogica\footnote{\,``...But if the matter was evenly disposed throughout an infinite space, it could never convene into one mass; but some of it would convene into one mass and some into another, so as to make an infinite number of great masses, scattered at great distances from one to another throughout all that infinite space.''} :\,\,``...Mas se a mat\'eria estava distribu\'{\i}da no espa\c{c}o infinito, ela nunca poderia se agregar numa \'unica massa; parte dela se aglutinaria numa massa aqui e outra acol\'a, de modo a formar um n\'umero infinito de grandes massas, espalhadas a grandes dist\^ancias uma das outras preenchendo todo espa\c{c}o infinito''. 

A visão unificada proporcionada pela gravita\c{c}\~ao universal, aliada com as leis da mecânica,  originou uma nova visão  do Universo.  O \textit{ Cosmos} seria representado por um modelo infinito, est\'atico e sem \textit{éter cósmico}\footnote{Na  tradição filosófica grega, o  éter (5o. elemento) seria a substância formadora dos corpos celestes.}, uma concep\c{c}\~ao  que influenciaria o pensamento cosmol\'ogico por mais de 200 anos (ver \textbf{Figura 2}). 

A física e astronomia posterior a Newton, incluindo a mecânica dos fluidos,  termodinâmica e teoria cinética dos gases,  n\~ao produziram um novo pensamento cosmol\'ogico. Nada substancial foi acrescentado a vis\~ao teórica newtoniana de um universo est\'atico, infinito, euclidiano e regido por sua auto-gravita\c{c}\~ao;  desde  William Herschel (1738-1822)\footnote{Herschel descobriu o planeta Urano e foi o maior  construtor de telesc\'opios do s\'eculo XVIII. Em 1789, ele publicou um cat\'alogo com mais de 1000 nebulosas \cite{C94}. Alguns desses  objetos eram observados a olho nu. Mas todos apareciam de forma difusa mesmo quando observadas pelos maiores telescópios da época. O  filósofo Imannuel Kant (1724-1804) sugeriu em 1755 que as nebulosas, tal como  a Via Láctea, seriam  ``Universos-Ilha'' formados por estrelas.} - o mais famoso astrônomo do s\'eculo XVIII -  até a emergência da cosmologia moderna em pleno século XX \cite{C94}. 




\section{Origens da Cosmologia Moderna}
As sementes da cosmologia moderna foram plantadas até meados da terceira década do século XX.  A nova visão do \textit{Cosmos}, distinta do legado newtoniano, foi também estimulada pelos mesmos ingredientes do tardio renascimento científico, a combinação de teoria, novas tecnologias e observações astronômicas. 

\begin{figure}
  \centering
 \includegraphics[width=3.4truein,height=2.8truein]{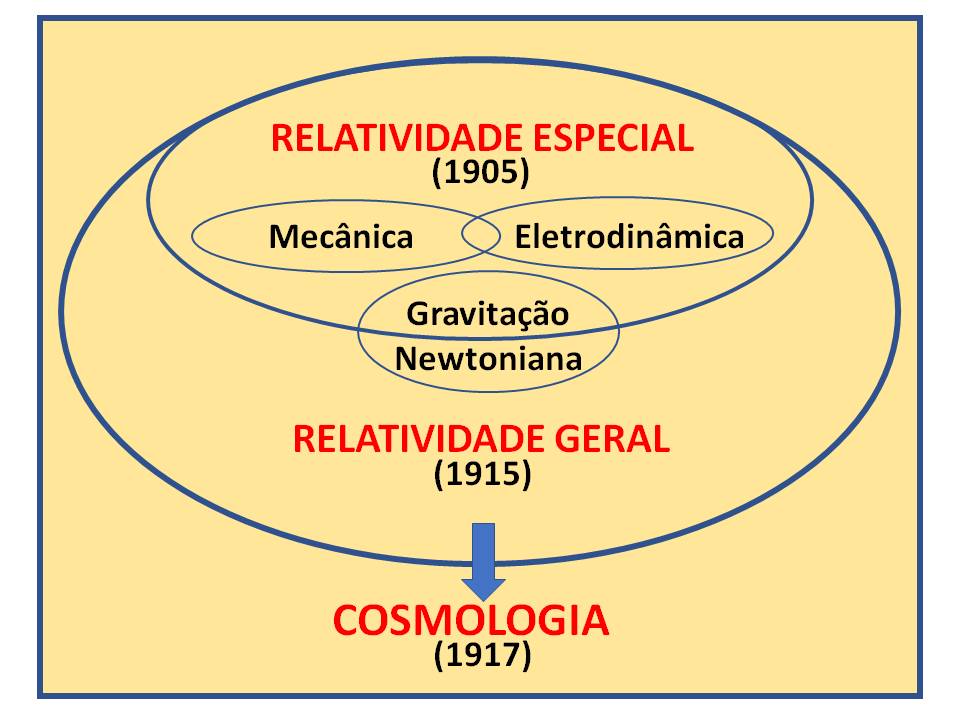}
\caption{Sementes teóricas da cosmologia moderna. A contradição entre  relatividade especial e gravitação newtoniana foi resolvida pela relatividade geral (1915).  Em 1917,  Einstein fundou uma nova área de pesquisa (Cosmologia), aplicando  sua teoria e tratando o Universo como um sistema físico.}\label{RERGCosm}
\end{figure}

Do lado teórico, as revolu\c{c}\~oes científicas marcando o início da f\'{i}sica moderna: mec\^{a}nica qu\^{a}ntica (1900), relatividade especial (1905) e relatividade geral (1915),  foram  decisivas \cite{AP82_0,Paty09}. 

Inicialmente, a relatividade especial (ou restrita) \cite{E05} revelou-se bem mais importante para a cosmologia do que a mecânica quântica\footnote{As idéias da microfísica e da teoria de radiação se revelariam inestimáveis a partir de 1965, com a descoberta da radiação cósmica de fundo (ver linha do tempo na \textbf{Figura 1}).}, pois requereu uma modifica\c{c}\~{a}o imediata da teoria gravitacional de Newton. De fato, a no\c{c}\~{a}o de a\c{c}\~{a}o \`a dist\^{a}ncia era  incompat\'{i}vel com as id\'eias de causalidade da relatividade restrita, segundo a qual os  campos f\'{i}sicos deveriam se propagar com velocidade finita. Em 1905, Henri Poincar\'e (1854-1912) já havia argumentado que uma descri\c{c}\~{a}o matem\'atica consistente do campo gravitacional deveria prever a exist\^{e}ncia de radia\c{c}\~{a}o gravitacional\cite{P05}.

Einstein começou sua busca por uma descrição relativística da gravitação em 1907 ao escrever seu primeiro artigo de revisão da relatividade \cite{E07}, provavelmente, estimulado pelos questionamentos de Poincaré. Em 1915, depois de muitas tentativas e erros durante os 3 últimos e mais intensos anos de sua busca\footnote{Uma descrição vívida e detalhada  desse período decisivo (1912-1915),  pode ser vista em \cite{AP82_0}.}, Einstein finalmente propôs a relatividade geral \cite{E15}. Em 1916, com a nova teoria ele demonstrou a conjectura de Poincaré\footnote{As ondas gravitacionais previstas por Einstein em 1916 foram observadas 100 anos depois \cite{PRL16}.}.  

Em 1917, Einstein  aplicou suas equa\c{c}\~{o}es para descrever o Universo, iniciando a cosmologia teórica do s\'eculo XX (ver seção 5).  Para obter  uma solução estática, Einstein modificou as equações de 1915, introduzindo a constante cosmológica $\Lambda$. Em  dois artigos subsequentes, Friedmann (1922, 1924) demonstrou que existiam soluções expansionistas (com e sem $\Lambda$), encerrando as contribuições teoricamente mais relevantes até o final da década. 

Na \textbf{Figura 3}, mostramos a conexão entre as sementes teóricas que originaram a cosmologia moderna. Diferente da gravitação newtoniana, o eletromagnetismo maxwelliano e a mecânica estavam contidos na relatividade especial\footnote{O eletromagnetismo revelou-se uma teoria relativística (sendo apenas despojado do conceito de éter luminífero), enquanto a mecânica foi imediatamente generalizada para satisfazer o critério relativístico especial de causalidade, $V\leq c$.}.  A contradição com a gravitação foi completamente resolvida pela relatividade geral e 2 anos depois Einstein inicia a cosmologia teórica moderna (1917), como um novo ramo da física.   

Do lado tecnológico/observacional, a cosmologia foi impulsionada pela constru\c{c}\~{a}o dos grandes telesc\'opios e também pelo aprendizado adquirido sobre medidas de velocidades e distâncias das nebulosas (como as galáxias eram chamadas na época). 

Em julho de 1917, entrou em operação o grande telescópio Hooker (refletor) de 2,5m de diâmetro, localizado no Monte Wilson (Califórnia), o mamute da época\footnote{Em celebração ao centenário da primeira luz do Hooker, foi disponibilizado pelo Observatório de Monte Wilson o histórico de instalação desse importante instrumento para o pioneiro estudo das galáxias \cite{MountWilson}}. Naquela altura ainda  persistia o debate entre astr\^{o}nomos para determinar o tamanho da Via L\'actea. O problema era saber se as nebulosas eram parte da nossa galáxia ou seriam outros ``Universos-Ilha'' como especulado por Kant (ver nota de rodapé 7).  Na segunda década do século XX, esta dúvida ainda não resolvida já tinha mais de 150 anos.  

A polêmica  estava relacionada com as dificuldades na determina\c{c}\~{a}o de dist\^{a}ncias em astronomia. Em 1920, essa controv\'ersia foi o tema escolhido para duas apresentações que ficariam conhecidas como o \textit{Grande Debate} entre os astr\^{o}nomos Harlow Shapley (1885-1972) e Heber Curtis (1872-1942).  O debate terminou inconclusivo, mas acelerou o desenvolvimento e o interesse pela imensidão do \textit{Cosmos}, uma área do conhecimento que no futuro próximo seria chamada de  astronomia extragal\'actica \cite{Hbook,Marcia2009,JL08,LL17}.


Por outro lado, medidas de velocidades radiais de afastamento (ou de aproximação) das nebulosas vinham sendo medidas desde 1914 pelo astrônomo americano Vesto Slipher (1875-1969). Tais medidas  são relativamente simples e de grande precisão, pois são baseadas no chamado efeito Doppler da luz emitida pelos corpos celestes. Para objetos que se aproximam (z negativo) ou se afastam (z positivo), tal efeito se traduz no desvio para o vermelho ou azul de suas linhas espectrais
\begin{equation}
z\equiv (\lambda_0 - \lambda)/\lambda = V/c\,,\,\,\,\,(V<<c)
\end{equation} 
onde  z é o parâmetro de \textit{ ``redshift''}, V é a velocidade da fonte e  $\lambda_0$,  $\lambda$  são, respectivamente,  os comprimentos de onda recebido pelo observador e emitido pela fonte.

Em 1923, uma tabela com 41 medidas, das quais 36 com \textit{redshifts} positivos ($z \geq 0.006$) e apenas 5 com valores negativos, incluindo Andromeda (M31) com $z=-0.001$,   foi publicada no prestigiado livro de relatividade geral\footnote{A tabela solicitada por Eddington foi preparada por Slipher e incluia seus últimos dados. Alguns não haviam sido publicados.} do astrofísico inglês Arthur Eddington (1882-1944) \cite{AEd23}. Contudo, na época não estava ainda claro o seu significado físico, pois não se sabia como estimar as distâncias para as nebulosas. 

Ainda em 1923, Edwin Hubble  (1889-1953),  utilizando o telescópio Hooker (ver \textbf{Figura 4}), descobriu estrelas variáveis Cefeídas em Andrômeda. Usando essas estrelas como vela-padrão\footnote{As pulsantes Cefeídas foram calibradas por Henrietta  Leavitt (1868-1921) permitindo medir dist\^{a}ncias extragal\'acticas, uma possibilidade levantada por Harlow Shapley em 1920, durante o Grande Debate.} Hubble determinou a dist\^ancia para a gal\'axia de Andr\^omeda (M31) estabelecendo que estava efetivamente fora da Via Láctea. Essa solução para o \textit{Grande Debate} marcou o início da astronomia extragaláctica \cite{H24}. Finalmente, a verdadeira cosmologia observacional dava seus primeiros passos. Em breve começaria o acoplamento entre teoria e observação. 

\begin{figure}
 \centering
\includegraphics[width=2.4truein,height=2.8truein]{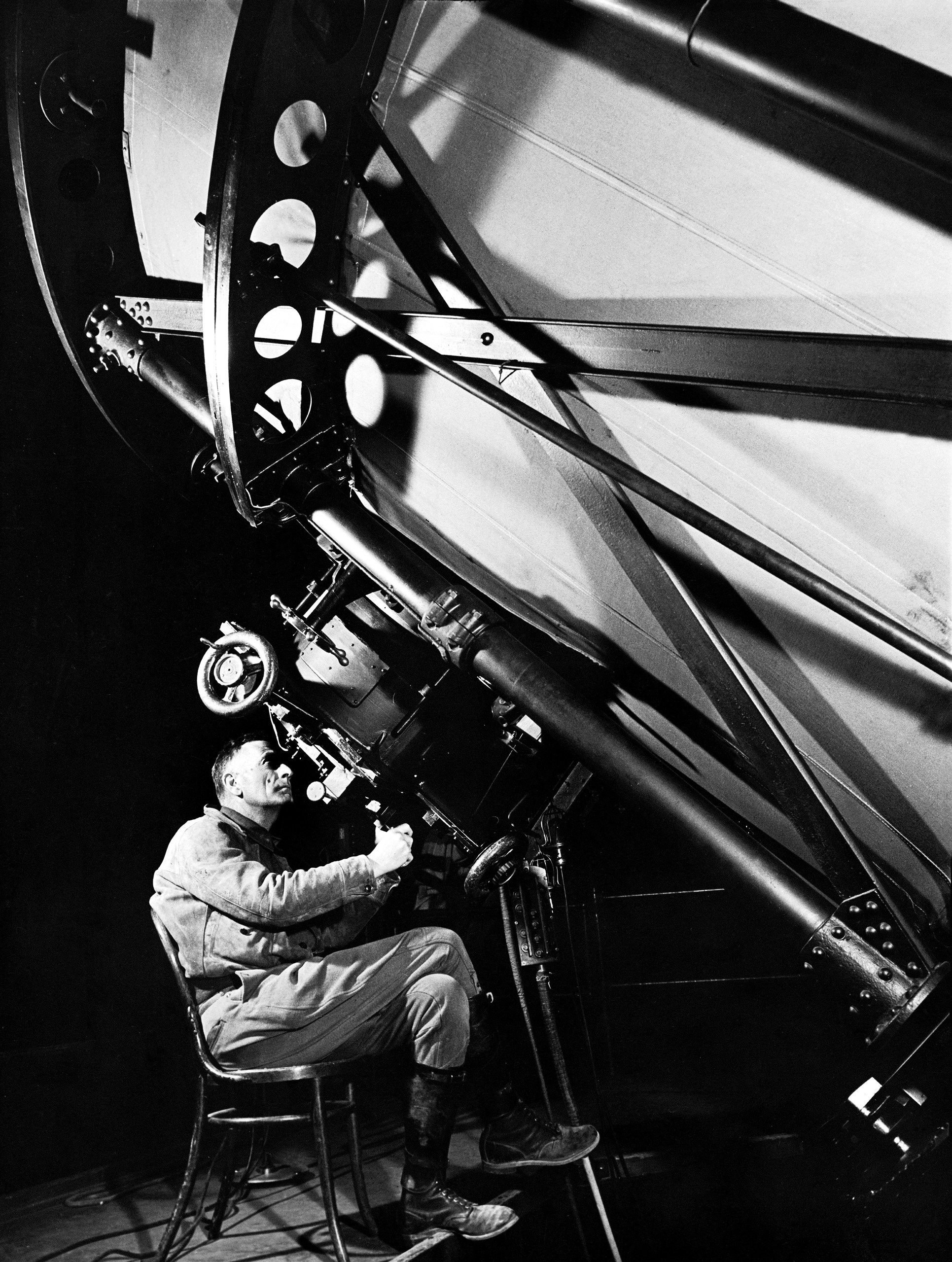}
\caption{Hubble no telescópio Hooker no Observatório do Monte Wilson (Califórnia). Crédito da imagem:\cite{HTel}.}
\label{FigN} 
\end{figure}

É importante enfatizar que do ponto de vista teórico ocorreram duas rápidas mudanças paradigmáticas  em relação as idéias newtonianas. De um lado, uma nova teoria de gravidade mais precisa do que a newtoniana  e de outro, um novo modelo cosmológico foi proposto, mas agora como solução matemática da relatividade geral.  Devido algumas nuances e sutilezas que precisam ser compreendidas, examinaremos os passos fundamentais da transição Relatividade Geral-Cosmologia nas duas seções a seguir. Como veremos, o resultado dessa sinergia entre teoria e observação será não apenas a previsão teórica, mas também a descoberta da expansão Universal.  
 
\section{Gravitação Einsteiniana: Relatividade Geral}

Em 1907, Einstein percebeu que a força gravitacional newtoniana era de certa forma ilusória, pois um observador em queda livre não sente seu próprio peso. A possibilidade de anular localmente a gravidade indo momentaneamente para o sistema em queda livre, significa que a interação gravitacional pode ser pensada como uma força fictícia.  Inversamente, um campo gravitacional estático e homogêneo, $\vec g$, pode também ser simulado por um referencial uniformemente acelerado com $\vec a= -\vec g$, um atributo válido na gravidade newtoniana.  

Em 1912, Einstein nomeou essa propriedade de \textit{Princípio de Equivalência} (PE). Na verdade,  Einstein entendera desde 1907 a existência de  uma equivalência física entre efeitos gravitacionais e acelerativos, cujas consequências precisavam ser investigadas\footnote{Após o artigo de 1907, Einstein retornaria ao problema gravitacional somente em meados de 1911.}. Contudo, apesar de algumas tentativas mal sucedidas, até meados de 1912, ainda não estava claro para Einstein como  essa equivalência seria utilizada para construir uma teoria relativística da gravitação. Einstein entrara no labirinto de teorias com velocidade da luz variável (devido a presença da gravitação), provavelmente, pela perda de um atalho precioso em sua busca; a noção de espaço-tempo métrico e suas conexões. 

Em 1908, Herman Minkowski (1864-1909)\footnote{Minkowski foi professor de Einstein na graduação do Instituto Federal de Tecnologia em Zurique (ETHZ).} havia proposto um conceito que seria muito importante para a relatividade geral; a noção de um contínuo espaço-tempo e sua descrição métrica \cite{M08}. Ele demonstrou que a propriedade que  distingue as transformações de Lorentz e de Galileo é a invariância da métrica (elemento de linha)  quadrimensional:
\begin{equation}\label{M1}
ds^{2}= c^{2}dt^2 - {dx^2} - dy^{2} - dz^{2}= \eta_{\alpha\beta}dx^{\alpha}dx^{\beta}\,,  
\end{equation}
onde $\eta_{\alpha\beta}$  é o tensor métrico ou matriz de Minkowski,  o 4-vetor $x^{\alpha}\equiv (ct,x,y,z)$ e os índices gregos $\alpha, \beta=0,1,2,3$. Note que na segunda igualdade foi adotada a convenção de Einstein - índices repetidos indicam soma. Da primeira igualdade lemos diretamente os coeficientes métricos $\eta_{00} = 1,\, \eta_{0i}=0, \, \eta_{ij}=-\delta_{ij}$ ($i,j =1,2,3$). Portanto, a matriz é definida apenas pelos elementos diagonais, $\eta_{\alpha\beta}$ $\equiv diagonal\, (+1, -1,-1,-1)$.  

Embora seja uma métrica pseudo-euclidiana\footnote{Diferente do caso  pseudo-euclideano, uma métrica euclidiana 4-dimensional tem diagonal (+1, +1, +1, +1).}, é possível mostrar que a curvatura riemanniana do espaço-tempo de Minkowski é nula, ou seja,  o espaço-tempo quadrimensional é plano. Podemos afirmar que Minkowski substituiu as duas métricas separadas da mecânica newtoniana (tempo absoluto e espaço absoluto), ambas invariantes por transformações de Galileo  
\begin{equation}\label{MG}
(I)\,\,\,dt^{2}=dt'^{2},\,\,\,\,\,(II)\,\,\,d\ell^{2} = {dx^2} + dy^{2} + dz^{2}=d\ell'^{2}\,,
\end{equation}
pelo contínuo 4-dimensional \eqref{M1}, absoluto,  plano e invariante por transformações de Lorentz (cf. \eqref{M1}). 

A nova linguagem matemática baseada em 4-vetores e 4-matrizes ou, mais geralmente, em tensores, simplificou consideravelmente a relatividade especial. Tal como aconteceu com a unidade entre espaço e tempo, foi também possível unificar os conceitos de energia e momentum, no 4-vetor de energia-momentum, $P^{\mu}\equiv (E/c, \vec p)$. Para um meio material contínuo ou um campo físico com densidades de energia e momentum, a unidade é estabelecida pelo 4-tensor de energia-momento, $T^{\mu\nu}$.  Na época, a radicalidade dessas contribuições de Minkowski não foi percebida por  Einstein\footnote{Ao ter conhecimento do artigo de Minkowski, Einstein o classificou como erudição supérflua \cite{AP82}. Anos depois mudaria de idéia.}.

Uma conexão anterior entre o formalismo de Minkowski e o princípio de equivalência teria abreviado sua busca,  como Einstein posteriormente reconheceria. De fato, se todos os referenciais são igualmente adequados para descrever a física, então a geometria euclidiana não pode ser válida em todos eles já que os referenciais acelerados afetam as réguas e o tic-tac dos relógios. Isto significa que os coeficientes da métrica do espaço-tempo,  não podem se reduzir em todos os lugares a forma de  Minkowski. Por outro lado, a equivalência entre efeitos acelerativos e gravitacionais sugeria claramente uma teoria geométrica para a gravitação baseada numa geometria não-euclidiana. Einstein encontrara as variáveis que descreviam o campo gravitacional, mas faltavam suas equações dinâmicas.  

Em 1913, inspirado por diferentes conceitos físicos e matemáticos,  incluindo: (i) o  Princípio de Equivalência, (ii) a idéia de espaço-tempo introduzido por Minkowski e, (iii) os estudos de Bernhard Riemann (1826-1866), Gregorio Ricci (1853-1925)  e outros geômetras sobre espaços curvos de dimensões arbitrárias, Einstein associou corretamente o efeito da gravitação a modificação da métrica do espaço-tempo Minkowskiano:    

\begin{equation}
ds^{2}= g_{\alpha\beta}dx^{\alpha}dx^{\beta}\,,  
\end{equation}
onde as componentes de $g_{\alpha\beta}$ são funções da posição e do tempo que em geral não podem ser globalmente reduzidas a forma de Minkowski dada por \eqref{M1}. Isso pode ocorrer apenas numa vizinhança espaço-temporal suficientemente pequena (para atender o PE), ou na ausência total de matéria, quando a solução global esperada seria a relatividade especial e a métrica de Minkowski, $g_{\alpha\beta}\equiv\eta_{\alpha\beta}$. Faltava apenas as equações satisfeitas pelo tensor métrico.   

Em 1915, Einstein finalmente propôs as equações descrevendo a interação gravitacional ou, equivalentemente, a geometria de um espaço-tempo curvo na relatividade geral \cite{E15}: 

\begin{equation}\label{EE} 
G_{\mu\nu} \equiv R_{\mu\nu} - \frac{1}{2}g_{\mu\nu} R = \chi T_{\mu\nu}\,,
\end{equation}
onde $\chi = 8\pi G/c^{4}$ é a constante de Einstein\footnote{A  presença da constante gravitacional de Newton (G) e da velocidade da luz (c) na constante de Einstein, significa que a Relatividade Geral é uma teoria de gravidade relativística.},  $R_{\mu\nu}$ é o chamado tensor de Ricci que descreve a geometria, $g_{\mu\nu}$ é o tensor  métrico do espaço-tempo, $R=g_{\mu\nu}R^{\mu\nu}$ é o escalar de curvatura de Ricci e $T_{\mu\nu}$ o tensor de energia-momentum, representando o conteúdo energético-material do Universo ($\mu,\nu=0$,1,2,3). 

Note que todas as quantidades  acima (tensores) são matrizes simétricas $4\times4$ ($G_{\mu\nu}=G_{\nu\mu}$,\,$T_{\mu\nu}=T_{\nu\mu}$). Isto significa que  no caso geral (sem simetrias) as equações acima reduzem-se a 10 equações diferenciais parciais acopladas independentes. 

\begin{figure}
\centering
 \includegraphics[width=3.4truein,height=2.9truein]{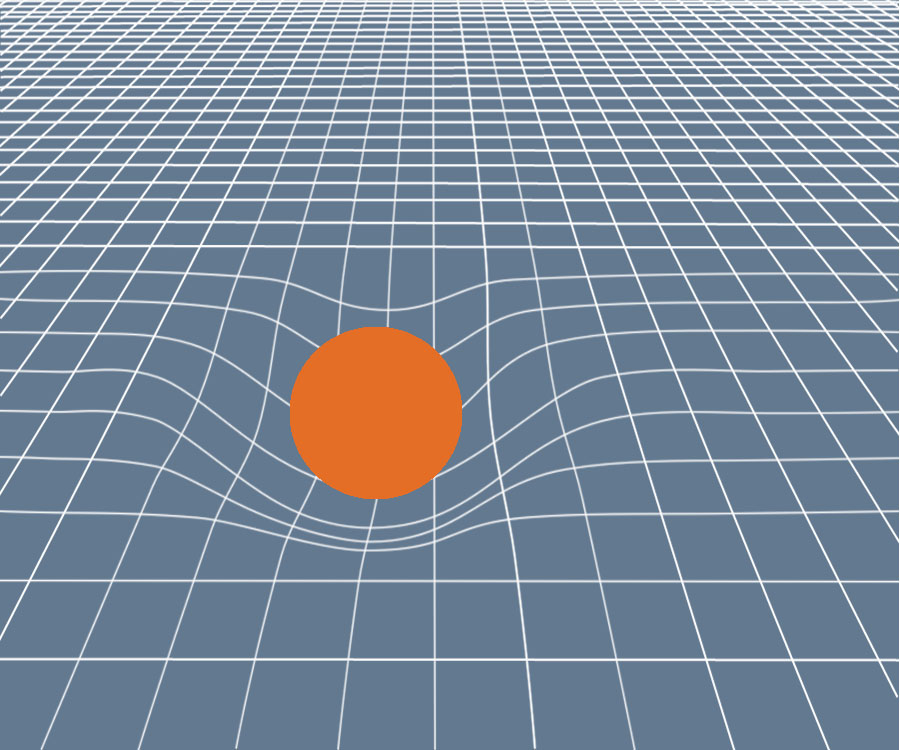}
\caption{Geometria,  curvatura e matéria.  De acordo com as equações \eqref{EE} propostas por Einstein, a geometria do espaço-tempo depende da presença da matéria. Numa situação estática como na figura,  vemos que a relatividade especial é válida longe das fontes. Neste caso, o contínuo espaço-tempo é descrito pela métrica \eqref{M1} descoberta por Minkowski.} \label{UnivNewton}
\end{figure}

Na \textbf{Figura 5} ilustramos pictoricamente a conexão entre geometria, espaço-tempo e curvatura numa situação simples, o caso de uma massa isolada. Qualitativamente, as equações de Einstein podem ser entendidas como um compromisso entre a curvatura (geometria) e o conteúdo material-energético do espaço-tempo representado pelo  tensor de energia-momentum. A geometria passa a ser um campo dinâmico determinado pela distribuição de matéria. Qualitativamente, as equações de Einstein \eqref{EE} podem ser representadas da seguinte forma:  

\[
\begin{bmatrix}
 Matriz\,\,\,descrevendo\\
 a\,\,\, geometria \,\,\, do \\  Espaço-Tempo
\end{bmatrix}
=
\begin{bmatrix}
     Matriz\,\,das\,\,fontes\\
	 \,\, matéria-energia\\do\,\, Espaço-Tempo 
\end{bmatrix}
\]
ou ainda, nas palavras de John Wheeler \cite{Wheeler}
\[
\begin{bmatrix}
  Geometria\,\, diz \,\, como \\ Matéria-Energia \\ devem\,\, 
  {\large \textbf{MOVER-SE}} 
\end{bmatrix}
=
\begin{bmatrix}
 Matéria - Energia \\  
 obriga\,\,\, Geometria\\ \,\, {\large\textbf{a\,\,\, CURVAR-SE}}
\end{bmatrix}
\]

A radicalidade da teoria de Einstein é precisamente o fato da geometria do espaço-tempo não ser mais um dado a priori como na física newtoniana ou mesmo na relatividade especial. O espaço-tempo em geral passa a ter uma dinâmica. 

É interessante que a cosmologia relativística foi iniciada  por Einstein antes mesmo da relatividade geral tornar-se uma teoria bem aceita ou conhecida. De fato, foi a experiência da deflexão da luz das estrelas pelo Sol - realizada no eclipse solar total em maio de 1919 - que disparou o processo de reconhecimento da teoria. 

O experimento foi realizado simultaneamente na Ilha do Príncipe (na costas da Guiné Equatorial) e na cidade de Sobral, no Ceará. A consagração da teoria (e do autor) ocorreu em 6 novembro do mesmo ano, numa reunião conjunta da Royal  e da  Astronomical Society em Londres.  Na ocasião, o físico J. J. Thompson fez o seguinte anúncio: "...a previsão e confirmação da deflexão da luz é o resultado mais importante relacionado com a teoria da gravitação desde os tempos de Newton e representa uma das maiores conquistas do pensamento humano" \cite{AP82_0,Lima99}. Podemos afirmar hoje que um dos maiores bônus intelectuais de Einstein pela descoberta da Relatividade Geral, foi o imediato lançamento da pedra fundamental da cosmologia moderna, como um novo ramo da física (\textbf{ver Figura 3}).  

\section{O Universo Estático de Einstein (1917)}
\textit{Why $\Lambda$ or not $\Lambda$ became the great doubt of cosmology?}\footnote{\textit{Por que  $ \Lambda$ ou não-$\Lambda $ tornou-se a grande dúvida da cosmologia?}}
{\vskip 0.3cm}
A falta de observa\c{c}\~{o}es do universo profundo, aliada as considerações cosmológicas de Newton sobre  um Universo infinito e estático, bem como as suas  dificuldades conceituais relativas as condições de fronteira no infinito,  induziram Einstein \cite{E17} a propor um modelo de Universo baseado em  tr\^{e}s hip\'oteses: 
 \vskip0.2cm
(\textbf{ i})  Mat\'eria c\'osmica descrita por um fluido perfeito sem press\~{a}o,
\vskip0.2cm
(\textbf{ ii}) ``Princ\'{i}pio Cosmol\'ogico" (homogeneidade e isotropia espacial), 
\vskip0.2cm
(\textbf{ iii}) Universo finito e est\'atico.  
\vskip0.2cm
A hip\'otese (i) significa que  Einstein optou por uma descrição hidrodinâmica. Na época, o universo inteiro era a Via Láctea  e como o gás de estrelas é muito  dilu\'{i}do, a matéria poderia se comportar como um fluido perfeito sem press\~{a}o. 

O postulado (ii) é de  origem  puramente geom\'etrica. Fisicamente, significa que todas as quantidades  observ\'aveis ser\~{a}o invariantes por transla\c{c}\~{a}o (homogeneidade) e rota\c{c}\~{a}o (isotropia). É uma hipótese simplificadora, pois a geometria espacial fica dotada da maior simetria poss\'{i}vel; restringindo assim o n\'umero de quantidades geométricas  para descrever o modelo.  

Um teorema da geometria diferencial garante que essa hip\'otese \'e compat\'{i}vel com 3 geometrias espaciais distintas, especificadas pelo par\^{a}metro de curvatura, $k= +1$ (fechado e finito), $k = -1$ (universo hiperb\'olico, infinito), $k=0$ (universo plano, infinito). 

Einstein adotou o caso fechado (esfera $S^{3}$). A escolha foi apresentada por ele como vantajosa pois resolvia o problema de condi\c{c}\~oes de contorno ou de  fronteira do Universo.


A  hip\'otese (iii) é sem dúvida a mais severa e restritiva delas, pois um Universo est\'atico n\~{a}o seria compat\'{i}vel com as equa\c{c}\~{o}es propostas por Einstein em 1915. A razão física é bem simples: um fluido autogravitante, sem pressão, estático, homogêneo e isotrópico, colapsaria totalmente numa escala de tempo finita. 

Para compatibilizar um modelo matemático com as id\'eias de um Universo est\'atico vigentes desde a antiguidade grega  e ainda sobreviventes até o século XX, Einstein modificou de forma ``ad hoc" sua teoria. Uma iniciativa nada trivial. 

Em 1917, um Einstein receoso escreveu numa carta para seu amigo Paul Ehrenfest: ``...\textit{fiz uma coisa no que se refere a gravitação que de algum modo me expõe ao perigo de ser internado num hospital de malucos''}\cite{AP82a}. 

Einstein surpreendia mais uma vez. Introduzindo um termo proporcional a uma constante ($\Lambda$), ele alterou o sistema original de equações \eqref{EE}  para  a seguinte forma:

\begin{equation}\label{EEL}
R_{\mu\nu} - \frac{1}{2}g_{\mu\nu} R - \Lambda g_{\mu\nu}= \chi T_{\mu\nu}\,,
\end{equation}
onde $\Lambda$ ficou conhecida como constante cosmológica\footnote{Na introdu\c{c}\~{a}o de $\Lambda$, Einstein foi muito influenciado pelas idéias de Ernest Mach (1838-1916) na sua crítica aos fundamentos da mecânica newtoniana. Segundo Mach, a inércia de um corpo seria um efeito coletivo da presença de todas as outras massas do Universo; uma idéia que poderia ser concretizada através de um modelo cosmológico finito\,\cite{P82}.}.
\begin{figure}
\centering
  \includegraphics[width=3.1truein,height=2.9truein]{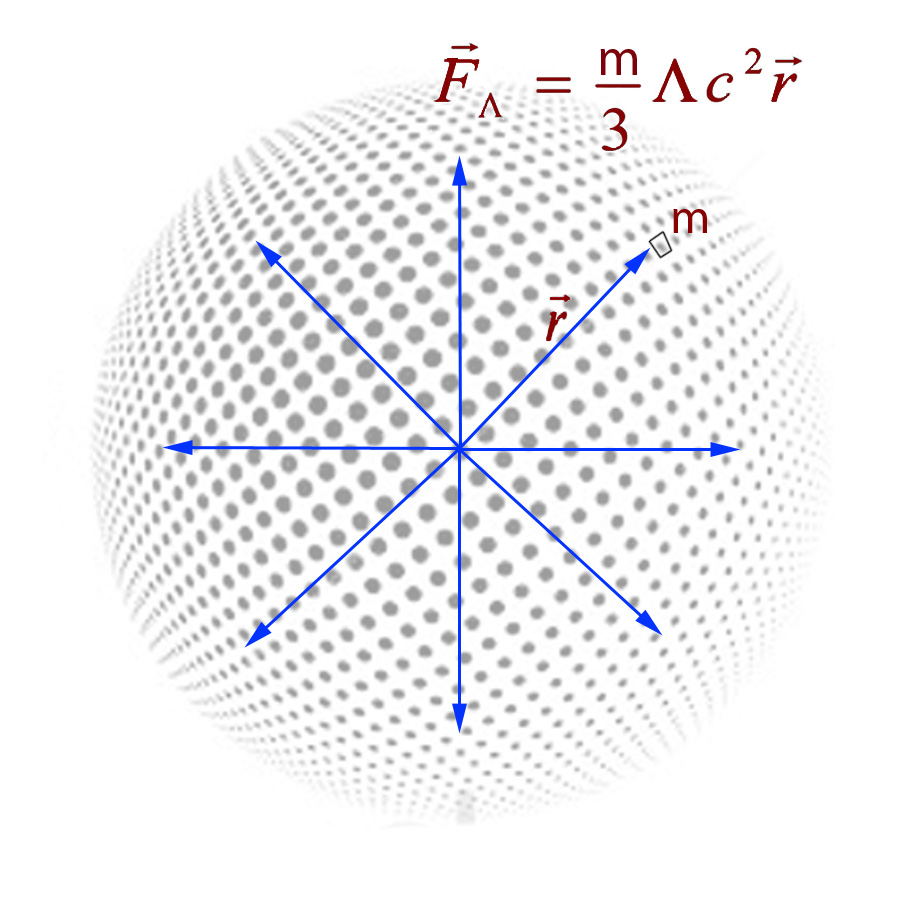}
\caption{Efeito de $\Lambda$ sobre uma massa $m$.  A constante cosmológica é equivalente a existência de uma força repulsiva proporcional a distância. A gravidade repulsiva de $\Lambda$ contrabalança o efeito da auto-gravitação da matéria evitando o colapso do modelo estático de Einstein.}\label{Lambda}
\end{figure}

O efeito físico de $\Lambda$ é muito fácil de entender. Na aproximação newtoniana, o campo gravitacional pode ser descrito pela famosa equa\c{c}\~{a}o de Poisson, ${\nabla^{2}}{\phi} = 4\pi G\rho$, onde $\phi$ \'e o potencial gravitacional (por unidade de massa) e $\rho$ a densidade de mat\'eria. Essa equação é também obtida de \eqref{EE}  na aproximação de campo fraco e baixas velocidades. Contudo, na modifica\c{c}\~{a}o de 1917, o limite é alterado para a seguinte expressão:  
\begin{equation}\label{Newt}
{\nabla^{2}}{\phi} = 4\pi G\rho - c^{2}\Lambda\,.  
\end{equation} 
Para uma distribuição esférica de matéria com densidade constante, a solução para $\phi$ a uma distância r (\textit {M} é a massa dentro do raio \textit{r}), é dada por:
\begin{equation}\label{phi}
{\phi} = -\frac{GM}{r} - \frac{c^{2}\Lambda}{6}r^{2}\,\,\,\Rightarrow \frac{F}{m}= -|{\vec\nabla} \phi|= -\frac{GM}{r^{2}} + \frac{1}{3}c^{2}\Lambda r. 
\end{equation}
Da segunda equação acima vemos que $\Lambda$ tem dimens\~{a}o de (comprimento)$^{-2}$ e gera um campo de for\c{c}a gravitacional repulsivo proporcional à distância\footnote{No seu  artigo original \cite{E17}, Einstein escreveu: ${\nabla^{2}}{\phi} - \Lambda\phi= 4\pi G\rho \,$, uma equação que não coincide com o limite de campo fraco e baixas velocidades de sua teoria com o termo $\Lambda$ \cite{Heck42,Alex00,BT88}.}. 

Na \textbf{Figura 6}, mostramos o efeito de $\Lambda$ provocando uma força  repulsiva proporcional a distância, $|\vec{F}|= m\Lambda c^{2}r/3$, sobre qualquer partícula de massa $m$. 
  
Por outro lado, é possível mostrar que as 10 equações de campo (\ref{EEL}) combinadas com as hipóteses (i)-(iii) reduzem-se trivialmente  para apenas duas equações independentes:
\begin{equation}\label{E1}
8\pi G\rho + \Lambda c^{2} = \frac{3c^{2}}{a^{2}}\,,
\end{equation}
\begin{equation}\label{E2}
\Lambda = \frac{1}{a^{2}}\,.
\end{equation}
 Na equação (\ref{E2}) acima $a$ \'e o raio do Universo espacialmente fechado (esfera $S^{3}$ de volume $V=2\pi^{2}a^{3}$).  Combinando (\ref{E1}) e (\ref{E2}) segue que $\rho=\Lambda c^{2}/4\pi G$. A constante cosmológica determina a densidade média de matéria e o raio do modelo estático. A massa total do modelo de Einstein é finita, $M= \rho\times 2\pi^{2}a^{3} = \pi c/4G\sqrt{\Lambda}$.

Tais resultados  demostram também que a constante cosmol\'ogica é desprez\'{i}vel na f\'{i}sica do sistema solar, mas pode ter import\^{a}ncia crucial nas grandes escalas cosmol\'ogicas. No modelo de Einstein, por exemplo, $\Lambda$  anula  o efeito atrativo da auto-gravita\c{c}\~{a}o do fluido,  evitando o colapso da mat\'eria c\'osmica e possibilitando uma solução estática e finita.

É interessante que o \textit{Cosmos} newtoniano homogêneo e infinito (\textbf{Figura 2}) é problemático devido ao fato do potencial gravitacional divergir no infinito, enquanto o  resultado exato de Einstein segue diretamente das hipóteses (i)-(iii) e da relatividade geral.  Observe que o modelo  obtido não  descreve uma aproximação como na discussão do significado de $\Lambda$, onde apelamos para uma aproximação de campo fraco da relatividade geral  (incluindo o termo $\Lambda$).

Nas conclusões do seu artigo, Einstein afirma que não discutirá se o modelo é defensável do ponto de vista observacional. Além disso, confessa  que o resultado foi obtido introduzindo uma extensão das equações de campo (o termo $\Lambda$) que não se justificava pelo conhecimento gravitacional da época\footnote{"In order to arrive to this consistent view, we admittedly had to introduce an extension of the field equations of gravitation which is not justified by our actual knowledge of gravitation."}. Finaliza dizendo que o termo $\Lambda$ é necessário apenas para tornar possível uma distribuição estática de matéria, como requerida pelas pequenas velocidades estelares. Nesse ponto vemos que Einstein não tinha conhecimento das medidas de velocidade das nebulosas. 

Em 1917, Willem de Sitter (1872-1934) obteve outra solução estática\footnote{Em 1929, Howard Robertson (1903-1961) mostrou que as soluções de Einstein e de Sitter eram as únicas soluções cosmológicas estáticas das equações de Einstein\cite{R29}.} das equações de Einstein tendo $\Lambda\neq 0$, mas sem matéria \cite{dS1}. Essa solução não segue das equações de Einstein \eqref{E1}-\eqref{E2} tomando o limite $\rho \rightarrow 0$, pois a geometria é completamente distinta.  Na solução de de Sitter, a curvatura escalar $R$ do espaço-tempo (escalar de Ricci)  é determinada por $\Lambda$. De fato, na ausência de matéria ($T_{\mu\nu}\equiv 0$), o traço da equação \eqref{EE} reduz-se para $R=4\Lambda$.

A solução de Einstein continha mat\'eria sem movimento, enquanto a de de Sitter permitia movimento sem mat\'eria. \textit{Como isso é possível tendo $\rho=0$\,?} Na verdade, toda a matéria observada, incluindo as nebulosas,  seriam partículas de teste na geometria de de Sitter e, sob ação de $\Lambda$,  adquirem uma velocidade de afastamento. Posteriormente, de Sitter e outros tentaram associar esse efeito ao desvio Doppler observado nas linhas espectrais, o que ficou conhecido como \textit{Efeito de Sitter}\footnote{A nova solução revelou-se também um duro golpe na relatividade da inércia que Einstein chamava de ``Princípio de Mach". Ao contrário do que pensava Einstein, $\Lambda$ não proibia uma inércia relativa ao espaço.}.

Einstein pagou um preço excessivo pela modificação de suas equações de campo.  A introdu\c{c}\~ao de $\Lambda$ custou-lhe  o que seria mais uma fant\'astica previs\~{a}o de sua lavra,  a recess\~{a}o das gal\'axias (numa vis\~{a}o mais newtoniana), ou a expans\~{a}o do Universo, segundo a relatividade geral. Podemos afirmar que uma conjectura científica bem articulada e psicologicamente enraizada na cultura (a idéia de Universo estático) afetou também um virtuose da física acostumado a fazer previs\~{o}es surpreendentes. Como veremos, a constante $\Lambda$ foi enterrada e ressuscitada diversas vezes nos últimos 100 anos. Alegoricamente, como nas histórias de Aladdin e o gênio,  podemos dizer que  $\Lambda$  havia saído da garrafa e recusava-se a entrar novamente.   

\section{Modelos Expansionistas do Universo (1922)}

Em 1922, o meteorologista e cosm\'ologo russo, Alexander Friedmann (1888-1925),  tornou-se o pai da teoria do ``Big-Bang''\footnote{A designação modelos de \textit{ Big-Bang}  para as soluções expansionistas foi sugerida por Fred Hoyle, numa entrevista a BBC de Londres em 1949. Era uma crítica, mas revelou-se um tremendo sucesso.}. Ele mostrou que as equa\c{c}\~{o}es de Einstein com curvatura positiva ($k=+1$) previam solu\c{c}\~{o}es expansionistas para um fluido sem press\~{a}o, tanto  na teoria de Einstein original (1915),  quanto na reformula\c{c}\~{a}o de 1917 incluindo $\Lambda$ \cite{F22}. Em 1924, Friedmann obteve novas solu\c{c}\~{o}es expansionistas (com e sem $\Lambda$) para o caso hiperb\'olico ($k = -1$) \cite{F24}. 

A métrica dos modelos expansionistas homogêneos e isotrópicos, ou seja,  cuja seção espacial ($t=const.$)  satisfaz o princípio cosmológico,  pode ser escrita como \cite{PBook}:

\begin{equation}\label{FRW}
ds^{2}= c^{2}dt^2 - a^2(t)\left[\frac{dr^2}{1 - k r^2}  +  r^{2}d\theta^{2} + r^{2}sen^{2}\theta d\phi^{2}\right],
\end{equation}
onde  ($r,\,\theta,\,\phi$) são coordenadas esféricas adimensionais (números puros). Essas coordenadas são ditas comóveis, pois os elementos de volume do fluido estão em repouso (ver \textbf{Figura 8}). O fator de escala $a(t)$ - a única quantidade  desconhecida na geometria - tem dimensão de comprimento e $k=\pm 1,0$ é o parâmetro de curvatura associado aos modelos fechado, aberto (hiperbólico) e plano, respectivamente\footnote{Na literatura moderna, o elemento de linha \eqref{FRW}  é bastante referido como métrica de Friedmann-Robertson-Walker (FRW). Em meados dos anos 30, Robertson \cite{R35} e e  Walker \cite{W36} mostraram que a métrica é independente da relatividade geral.}. Para $k=+1$ e $a=constante$, a métrica acima se reduz a forma adotada por Einstein no seu modelo estático.  

No caso de modelos fechados ($k=+1$),  o fator de escala é o próprio  raio (variável) do Universo, cuja seção espacial finita é representada pela esfera $S^{3}$ de volume $ V(t)=2\pi^{2}a^{3}(t)$.  Obviamente tal interpretação não é válida para os outros casos  ($k=-1,0$),  pois são espaços de volumes infinitos e com topologias (propriedades globais do espaço 3-dimensional) completamente diferentes da esfera $S^{3}$.  No entanto,  independente da curvatura, cada comprimento espacial finito ainda satisfaz a lei de escala, $L\propto a(t)$, ou seja, um volume finito qualquer varia satisfazendo,  $V(t) \propto a^{3}(t)$. 

Essa geometria deve ser compatibilizada via equações \eqref{EEL}  com uma fonte de curvatura. Para um fluido perfeito com densidade de energia, $\epsilon = \rho c^{2}$ e sem pressão $p=0$ (o mesmo fluido adotado por Einstein), o tensor de energia-momentum  tem forma,  $T_{\mu\nu}=\rho c^{2} u_{\mu} u_{\nu}$, onde  $u_{\mu}=\delta_{\mu\,0 }\equiv(1,0,0,0)$,  as equações de Einstein se escrevem:  


\begin{equation}\label{F1}
8\pi G \rho + {\Lambda}c^{2} = 3\frac{{\dot a}^{2}}{a^{2}} + 3\frac{kc^{2}}{a^{2}}\,,
\end{equation}

\begin{equation}\label{F2}
\Lambda c^{2} =   2\frac{\ddot a}{a} + \frac{{\dot a}^{2}}{a^{2}} + \frac{kc^{2}}{a^{2}}\,,
\end{equation}
\noindent onde um ponto denota derivada em relação ao tempo e $a(t)$ é o fator de escala .

É instrutivo verificar como o modelo estático de Einstein  é prontamente recuperado das equações acima. Considerando $\dot a = \ddot a = 0$,  segue  diretamente de (\ref{F2}) que $\Lambda = k/a^{2}$ e substituindo em  (\ref{F1}), obtemos $4\pi G\rho = kc^{2}/a^{2}$; o que permite uma solução com densidade de matéria positiva somente se $k=+1$. Nessas condições  o sistema (\ref{F1})-(\ref{F2}) se reduz ao sistema (\ref{E1})-(\ref{E2}). As equações mais gerais de Friedmann  não apenas recuperam o  resultado esperado,  $\Lambda = 1/a^{2}$, mas também explicam  a escolha geométrica de Einstein\footnote{Note que no caso estático os valores $k=0,-1$ não permitem soluções fisicamente consistentes ($\rho \leq 0$). Algo impossível de se perceber da leitura do artigo de Einstein!}. 

Friedmann obteve as soluções gerais das equações acima considerando $k=\pm 1$ e também $\Lambda=0$ e $\Lambda\neq 0$. A quantidade de matéria total dos modelos é constante, sendo finita no modelo fechado e infinita no modelo aberto. A densidade de matéria  diminui com a expansão, $\rho \propto a^{-3}(t)$, ou equivalentemente,  $\rho=\rho_0 (a_0/a)^{3}$, onde $\rho_0$ e $a_0$ são a densidade de matéria e o fator de escala hoje\footnote{Daqui em diante, o valor de uma quantidade qualquer com sub-índice zero, $X_0$, denotará o valor da quantidade $X(t)$  em $t=t_0$, seu valor atual.}.

Nos modelo de Einstein  $\Lambda$ determinava a densidade de matéria e o raio do Universo. Contudo,  Friedmann percebeu que $\Lambda$ nos modelos expansionistas é simplesmente um parâmetro livre, ou seja, uma constante arbitrária e que as soluções são extremamente simples no caso $\Lambda =0$. No caso geral pode-se mostrar que as soluções são funções elípticas, do tipo $t(a)$ que não podem ser invertidas para obter $a(t)$.  

Einstein relutou em aceitar as soluções expansionistas de Friedmann, chegando mesmo a publicar um curto artigo em 1922, alegando a existência de um erro matemático \cite{E22}. Friedmann escreveu para Einstein demonstrando que seu artigo estava correto, mas sem receber resposta, decidiu enviar também um emissário de nome Krutkoff que refez pessoalmente os cálculos com Einstein.  

Convencido, Einstein reagiu de imediato,  publicando na mesma revista outro pequeno artigo reconhecendo seu erro \cite{E23}. Afirmou também que as soluções de Friedmann estavam corretas, eram esclarecedoras e que de fato existiam soluções expansionistas espacialmente homogêneas e isotrópicas permitidas por suas equações de campo\footnote{\,``I am convinced that Mr. Friedmann's results are both correct and claryfying. They show that in addition to the static solutions to the field equations there are time varying solutions with a spatially symmetric structure''\,(Einstein \cite{E23}).}.

\begin{figure}
  \centering
  \includegraphics[width=3.1truein,height=2.5truein]{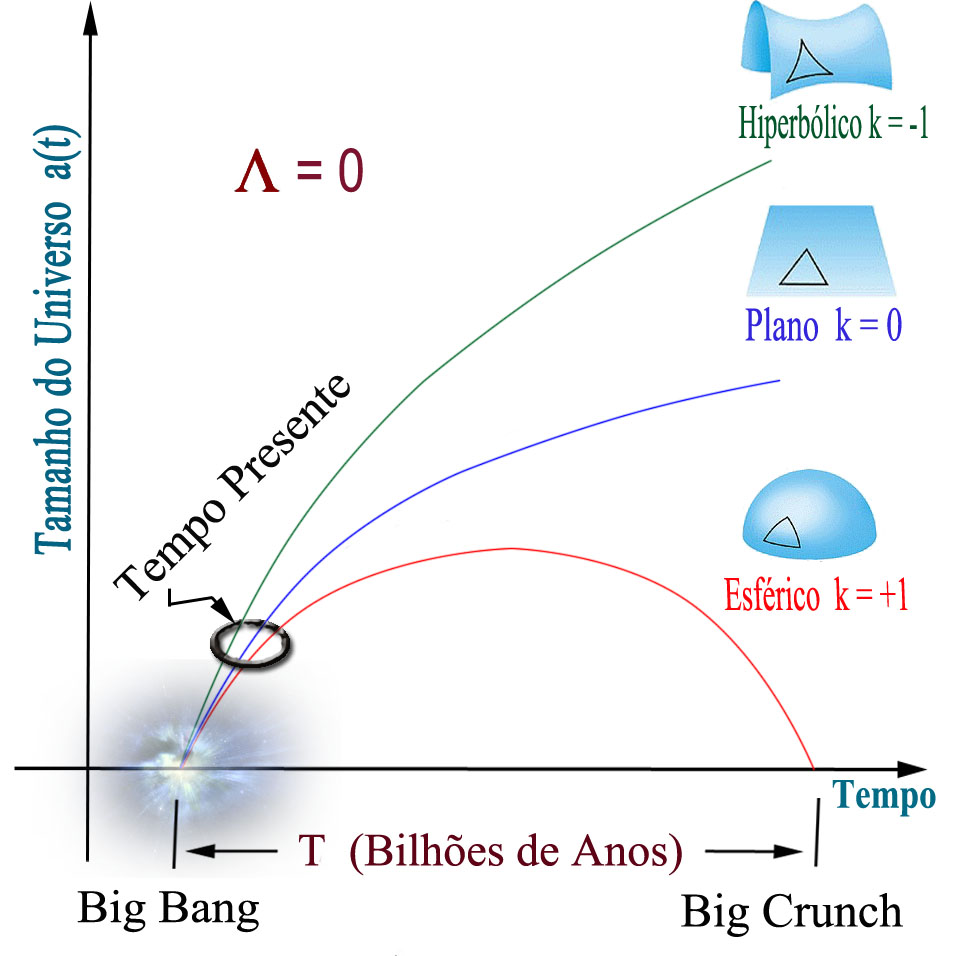}
\caption{Expansão do Universo em modelos do tipo Friedmann com $\Lambda=0$. A solução do caso plano ($k=0$) foi determinada por Einstein e de Sitter em 1932 (ver texto). Ao lado vemos os  análogos em 2 dimensões das geometrias espaciais 3-dimensionais, homogêneas e isotrópicas.}\label{Modelos}
\end{figure}

Faltava ainda as soluções expansionistas mais simples, a do caso plano ($k=0$). Considerando $\Lambda=k=0$,  o leitor pode verificar da  equação \eqref{F2} que a solução do fator de escala neste caso pode ser escrita na seguinte forma:
\begin{equation}\label{ES}
a(t) = a_0\left(\frac{t}{t_0}\right)^{2/3}\,,
\end{equation}
onde $t_0$ é a idade do Universo medida hoje, ou seja, o tempo total de expansão decorrido desde a singularidade ($t=0$).    

A solução acima foi obtido por Einstein e de Sitter\footnote{Einstein e de Sitter seguiram explicitamente a sugestão do astrofísico Otto Heckmann (1901-1983).} em 1932 \cite{ES32}, quando Einstein já havia renegado a constante cosmológica\footnote{A primeira negação (em privado) ocorreu no cartão postal de 23/05/1923 para o matemático e cosmólogo Herman Weyl (1885-1955). Os autores acreditam que o abandono foi uma decorrência natural do conhecimento de Einstein do modelo de Friedmann (1922), sem o qual não faria sentido eliminar $\Lambda$. Einstein escreveu: "Se não existe um mundo quase estático, então mandemos embora o termo cosmológico"\cite{Weyl}.}, completando o que ficaria conhecido na literatura como a classe de solu\c{c}\~oes cosmol\'ogicas do tipo Friedmann\footnote{Friedmann faleceu em 1925 na cidade de Kiev,  v\'itima de um surto de febre tifo que grassou na R\'ussia p\'os-revolucion\'aria. Provavelmente, foi esse tr\'agico acontecimento que lhe impediu de obter a solu\c{c}\~ao mais simples ($k=0$), derivada por Einstein e de Sitter 7 anos depois.}.

Na \textbf{Figura 7}, apresentamos qualitativamente a evolução temporal do fator de escala para os três tipos de modelos ($k=0, \pm 1$), considerando $\Lambda=0$. Os modelos expandem  a partir da singularidade em $t=0$ (``Big-Bang").  No caso fechado, os modelos são cíclicos, evoluindo entre a grande explosão e a grande implosão (``Big-Crunch"). Além disso, fixando a massa do Universo,  $M=5\times 10^{21}$ massas solares, Friedmann estimou um valor para o período T da ordem de 10 bilhões de anos. 

É importante enfatizar que para os três modelos não existe um centro de expansão. 
A homogeneidade e isotropia espacial (princípio cosmológico) garantem a mesma visão típica de qualquer galáxia. Em outras palavras, as galáxias estão paradas, cada uma na sua própria coordenada, mas o espaço entre elas expande continuamente. O efeito resultante é uma recessão galáctica em qualquer linha de visada. 

Na \textbf{Figura 8}, mostramos para o caso fechado uma analogia bidimensional desse fenômeno de expansão sem centro. Tudo funciona como se galáxias fossem pontos pintados nas coordenadas  ($\theta, \phi$) fixas na superfície de uma bexiga sendo inflada (são  análogas as coordenadas comóveis da métrica do tipo FRW).  A área da esfera simula o espaço curvo, real, disponível. Note que os observadores de qualquer galáxia compartilham a mesma visão típica do Universo. Qualquer galáxia pode ser considerada o centro\footnote{As galáxias efetivamente não aumentam de tamanho, pois são inomogeneidades locais de matéria cuja densidade é várias ordens de grandeza maior do que a densidade média do Universo.}. Essa analogia bidimensional da expansão sem centro é válida para todos os valores de k na \textbf{Figura 7}.  A  diferença é que no modelo fechado ($k=+1$),  o raio do Universo vai efetivamente a zero no instante $t=0$, enquanto nos outros casos dizemos que o fator de escala $a(t)$ vai a zero; o que também, efetivamente, colapsa as distâncias físicas no ``Big-Bang''. Na verdade, a visão de uma expansão sem centro continua válida até mesmo quando incluímos $\Lambda$.

Neste ponto é conveniente introduzir os parâmetros cinemáticos observáveis que emergem das equações de Friedmann: $H$ (parâmetro de Hubble) e $q$ (parâmetro de desaceleração):


\begin{equation}\label{P3}
H(t) = \frac{\dot a}{a},\,\,\,\,\, q(t)= -\frac{a\ddot a}{{\dot a}^{2}},\,\,\,\,\, 
\end{equation}
$H(t)$ tem dimensão de inverso de tempo e mede a taxa de expansão do volume [${\dot V}/V= 3H(t)$], $q(t)$ mensura a variação da taxa de expansão, ou seja, se o universo acelera ($\ddot a > 0$) ou  desacelera  ($\ddot a <0$)\footnote{Como a gravidade é atrativa, a crença de que a expansão seria desacelerada era tão forte que os astrônomos introduziram o sinal negativo na definição do parâmetro de desaceleração, de forma que  $\ddot a <0$ implicaria $q>0$.}

\begin{figure}
  \centering
 \includegraphics[width=3.4truein,height=2.5truein]{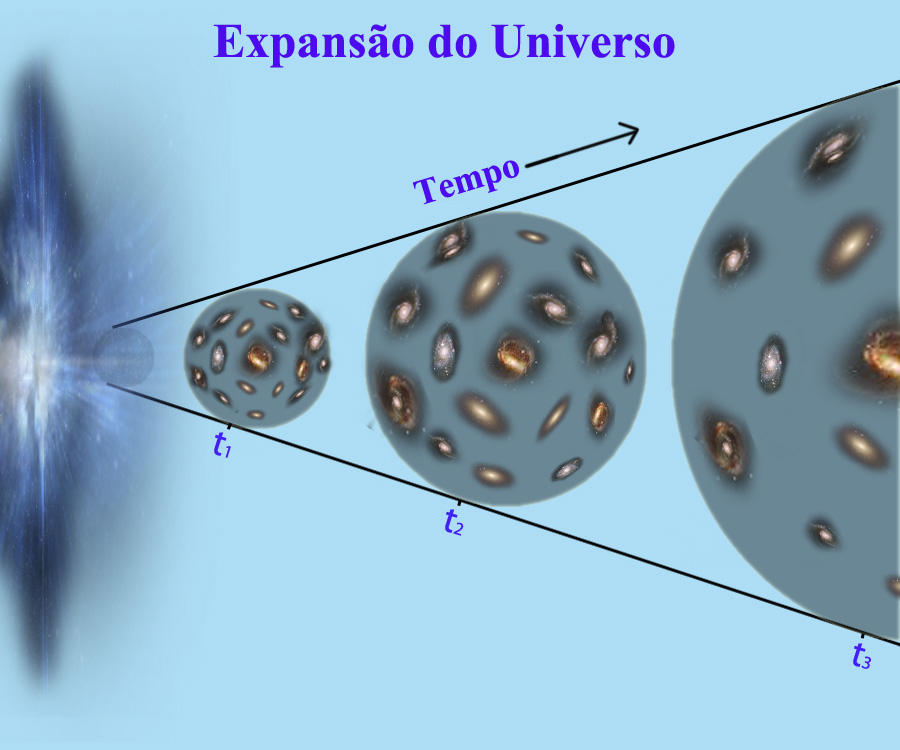}
\caption{Fenômeno da expansão sem centro. As coordenadas das galáxias pintadas na superfície do balão são fixas, mas o espaço entre elas expande aumentando a distância relativa.  Um observador em qualquer galáxia dirá que as outras se afastam, pois observa um \textit{redshift}  (recessão cósmica).}\label{ModeloExpansao}
\end{figure}

Observacionalmente,  os astrônomos determinam  os valores dessas quantidades hoje ($H_0$, $q_0$)  e via equações de Friedmann, podemos propagá-las  para o passado e também para o futuro \cite{X1, AL2011}.

Para exemplificar, considere as equações de Friedmann \eqref{F1}-\eqref{F2} para $\Lambda =0$. Neste caso é fácil deduzir a seguinte relação:

\begin{equation}\label{P4}
\frac{\ddot a}{a}= -\frac{4\pi G}{3}\rho\,, 
\end{equation}
válida para todos os valores de $k=0,\pm 1$ . Note que a expansão será sempre desacelerada (${\ddot a} < 0$), ou  seja, $q>0$, pois a densidade de matéria é positiva. Isso também explica a introdução (desnecessária) do sinal negativo na definição do parâmetro de desaceleração. 

Os artigos de Friedmann já haviam consolidado a possibilidade de um modelo dinâmico no âmbito da Relatividade Geral, mas não existia ainda qualquer sugestão de comprovação observacional. Essa incompleteza  teórico/observacional não duraria muito tempo.

\section{Descoberta da Expansão Universal (1929)}

Desde os anos 20, o Grande Debate demonstrara o interesse pela distância das nebulosas. As potencialidades do telescópio Hooker (ver nota de rodapé 12), também garantira um rápido desenvolvimento da recém-descoberta astronomia extragaláctica. Como resultado, os astrônomos finalmente definiram um método mais confiável para a  determinação de distâncias em cosmologia. 



Nosso  Universo era realmente bem maior do que o imaginado quando Einstein abriu a trilha teórica  da cosmologia moderna.   

A sinergia nessa combinação entre teoria, tecnologia e  observações astronômicas,  provocaria uma profunda mudança de paradigma. O uso intensivo do telescópio Hooker por Hubble, o transformaria numa espécie de Galileo da astronomia do século XX. Um campo inteiramente novo de pesquisa observacional tinha sido descoberto, lançando a humanidade na última fronteira astronômica, a pesquisa na vastidão do \textit{ Cosmos}. Uma vívida exposição, envolvendo o período até 1936, pode ser vista no  livro de Hubble, entitulado \textit {``O Domínio das Nebulosas"} (The Realm of the Nebulae) \cite{Hbook}.

Em 1929, Hubble mediu  dist\^{a}ncias at\'e as nebulosas pr\'oximas, utilizando  um número maior de estrelas Cefe\'{i}das como vela-padrão. Dois importantes resultados foram obtidos: (i) Confirmou-se para uma amostra maior (22 galáxias) que de fato as ``nebulosas''  n\~{a}o pertencem a Via L\'actea, resolvendo definitivamente  o debate sobre as ``nebulosas'' e o tamanho da Via Láctea; (ii) As gal\'axias est\~{a}o se afastando com uma velocidade proporcional a dist\^{a}ncia, ou seja exibem ``redshift'' ou  desvio para o vermelho das linhas espectrais.

Na \textbf{Figura 9}, apresentamos uma versão adaptada  do diagrama original de Hubble \cite{H29}.  A expansão universal prevista primeiramente por Friedmann em 1922  estava observacionalmente confirmada. Os dados revelaram um valor muito alto da constante na relação linear velocidade-distância, $H_0 = 536\, Km.s^{-1}. Mpc^{-1}$. Essa constante tem dimensão de (tempo)$^{-1}$, sendo hoje conhecida como constante de Hubble. 
                                                            
É extremamente  simples mostrar que a lei de Hubble pode ser obtida da métrica de Friedmann como uma aproximação de baixos \textit{redshifts} \cite{W2008}. Expandindo a(t) em torno de $t_0$ temos:

\begin{equation}\label{at}
a(t) \simeq a(t_0)[1 + H_0(t-t_0) +...],\,\,\,\,\, H_0={\dot a}(t_0)/a(t_0).
\end{equation}
Por outro lado, a razão dos fatores de escala em função do  \textit{redshift}  é  $a(t) = a_0/1+z$ (ver apêndice A). Na mesma aproximação podemos escrever a lei de Hubble, 
\begin{equation}
cz \simeq V = H_0c(t_0-t) + ...=H_0 d,
\end{equation}
onde aproximamos, $d\equiv c(t_0-t)$, para distâncias próximas. 
 
A descoberta da expansão representou uma profunda quebra de paradigma do Universo estático, uma visão aceita até 1922. Para os cosmólogos nunca houve dúvidas de que a descoberta de Hubble de fato representava a expansão do espaço (ver seção 8). 

Nas últimas linhas do seu artigo, Hubble  enfatizou  que a relação linear obtida é apenas uma primeira aproximação, pois as nebulosas {imageadas} estavam relativamente próximas; os dados obtidos se referem a um intervalo muito restrito de distância\footnote{``...It may be emphasized that the linear relation found in the present discussion is a first approximation representing a restricted range in distance''\cite{H1}.}.

\begin{figure}
  \centering
  \includegraphics[width=3.8truein,height=2.9truein]{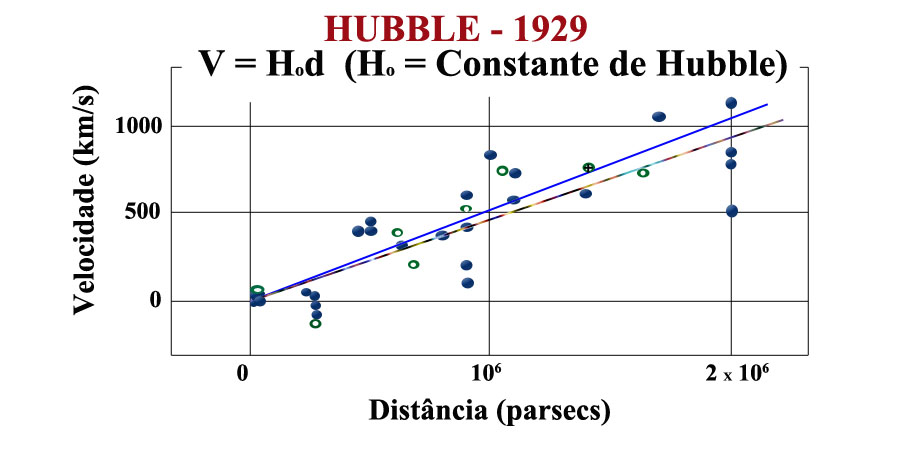}
\caption{Lei de Hubble.  Em 1929 Hubble utilizou medidas de redshift $z=V/c$ e de distância para 24 galáxias (discos pretos). A linha sólida é o melhor ajuste. A linha tracejada foi obtida para os mesmo dados binados (círculos). A cruz representa a velocidade média correspondendo a distância média de 22 galáxias cujas distâncias não puderam ser individulamente estimadas. Os dois conjuntos de dados sugerem um grosseiro comportamento linear entre a velocidade e a distância que ficaria conhecida como a Lei de Hubble.}
\label{Fig5} 
\end{figure}

Embora não tendo citado o trabalho de Friedmann, Hubble se referiu a cosmologia obtida por de Sitter (também sem citar o trabalho!), dizendo que a relação distância-velocidade poderia representar o chamado \textit{efeito de Sitter} (seção 5). Concluiu com um comentário ainda mais interessante, ao afirmar que dados númericos devem ser considerados nas discussões relativas a curvatura do espaço\footnote{``The outstanding feature, however, is the possibility that the velocity-distance relation may represent the de Sitter effect, and hence that numerical data may be introduced into discussions of the general curvature of space".}. Hubble estava consciente da relação entre o seu trabalho e as idéias relativísticas de espaço-tempo curvo mas, provavelmente, desconhecia os artigos de Friedmann, publicados em alemão.   
   
O valor de  $H_0$ medido por Hubble gerou um problema teórico para a cosmologia, o chamado problema da idade do Universo.  

Combinando \eqref{P3} com a solução \eqref{ES} segue que $H(t) = 2/3t$ e, portanto, a idade do Universo é dada por  $t_0=2H_0^{-1}/3$. Isto implica que o fator de escala na solução de Einstein-de Sitter \eqref{ES} pode ser escrito como (ver \cite{Lima01} para soluções gerais para fluidos perfeitos com pressão):
\begin{equation}
a(t)=a_0(\frac{3}{2}H_0t)^{2/3}\,. 
\end{equation}
Vemos que o inverso de $H_0$ determina o tempo de expansão desde o ``Big-Bang". No entanto, o valor obtido por Hubble era muito alto,  resultando numa idade do Universo muito baixa, em torno de $1,5$ bilhão de anos para o modelo plano, ou seja,  menor do que a idade da terra (a idade do modelo fechado era menor ainda e a do modelo hiperbólico um pouco maior). Como veremos a seguir, tal problema pode ser resolvido pela constante cosmológica.

\section{Sobre os Contributos de Lemaître}

Em 1927, Georges Lemaître (1894-1966)  publicou um artigo importante no pouco divulgado Anais da Sociedade Científica de Bruxelas \cite{L27}. O artigo foi publicado em francês e, por razões óbvias, não repercutiu na pequena comunidade da época trabalhando em cosmologia teórica e/ou astrofísica extragaláctica. Sua influência começou apenas em 1931, quando o editor da Monthly Notices (MNRAS) solicitou a Lemaître uma versão do artigo para publicação. Lemaître prontamente concordou e ele próprio traduziu para o inglês \cite{L31}, cortando  partes substanciais que julgou sem interesse devido ao trabalho de Hubble (1929).   

O artigo tem 4 resultados interessantes: (i) A conexão entre o desvio para o vermelho das linhas espectrais (redshift)  e a expansão do Universo, (ii) Uma dedução da lei de Hubble para pequenos redshits, (iii) alguns possíveis efeitos físicos de uma segunda componente cósmica, um fluido radiativo em equilíbrio termodinâmico de pressão $p=\rho/3$, e (iv) uma estimativa de $H_0 = 625 km/s/Mpc$, da mesma ordem de grandeza do valor de Hubble\footnote{No entanto, diferente do resultado de Hubble, o seu valor foi obtido como uma distância média estimada das nebulosas para exibir o valor de $H_0$.}. 

O artigo não foi citado por Hubble e, provavelmente, por nenhum autor entre 1927 e 1929, incluindo o artigo de Robertson sobre os fundamentos da cosmologia \cite{R29}.  Certamente teria repercutido antes do artigo de Hubble, caso tivesse sido publicado numa revista tradicional. Apesar disso,  a tradução com  supressão de seções efetuada pelo próprio Lemaître,  na prestigiada Monthly Notices (MNRAS)\footnote{Este fato foi recentemente esclarecido pelo astrônomo Mario Lívio \cite{ML11}, ao consultar os arquivos da MNRAS.} em 1931,  contribuiu para uma aceitação maior da expansão do espaço e, portanto, para o modelo relativístico de \textit{big-bang} tornar-se o modelo padrão da cosmologia\footnote{Como veremos na \textit{Parte II}, o modelo de Einstein-de Sitter ($k=0$) foi aos poucos se estabelecendo como modelo cosmológico padrão desacelerado para as idades média e moderna da cosmologia relativística (ver \textbf{Figura 1})}. 

Por outro lado, existia o problema da idade. Uma solução possível seria um erro no valor estimado por Hubble para $H_0$, talvez devido a um problema de calibração nas Cefeídas. Um valor menor de $H_0$ aumentaria a idade do Universo,  o que de fato comprovou-se nas gerações seguintes (ver Partes II). Outra possibilidade seria considerar uma solução teórica, tal como  sugerida por  Lemaître ao apelar para a constante cosmológica.

Em 1931, Einstein já havia abjurado definitivamente a constante cosmológica, a ponto de não considerá-la na sua solução de 1932 para o caso plano em colaboração com de Sitter (cf. seção 6 e nota de rodapé \textit{32}).

No entanto, Lemaître argumentou que a constante cosmológica (sendo uma força repulsiva), acelerava o Universo e aumentava a idade, potencialmente, resolvendo a contradição com o alto valor de $H_0$. O argumento é extremamente simples. Combinando as equações de Friedmann \eqref{F1}-\eqref{F2} é fácil deduzir a seguinte relação:

\begin{equation}\label{P4}
\frac{\ddot a}{a}= -\frac{4\pi G}{3}\left(\rho -\frac{\Lambda}{4\pi G}\right)\,. 
\end{equation}
Note que para $\Lambda=0$ a expressão acima  reduz-se a relação \eqref{P4} dos modelos desacelerados (${\ddot a} < 0$). No entanto, se $\Lambda >0$, como a densidade de matéria diminui com o tempo, o modelo pode acelerar quando $\rho (t) < \Lambda/ 4\pi G$, ou seja, quando $q<0$.

Modelos acelerados tem uma idade maior e a razão é bem simples. Todos os modelos (acelerados ou não) devem chegar ao presente valor do fator de escala. Isso significa que modelos acelerados partem de um instante anterior e, portanto, gastam um tempo de expansão maior para atingir o estágio atual. Uma prova baseada no cálculo explícito da idade do Universo será apresentada na \textit{Parte III}.

\begin{figure}
  \centering
 \includegraphics[width=3.4truein,height=2.6truein]{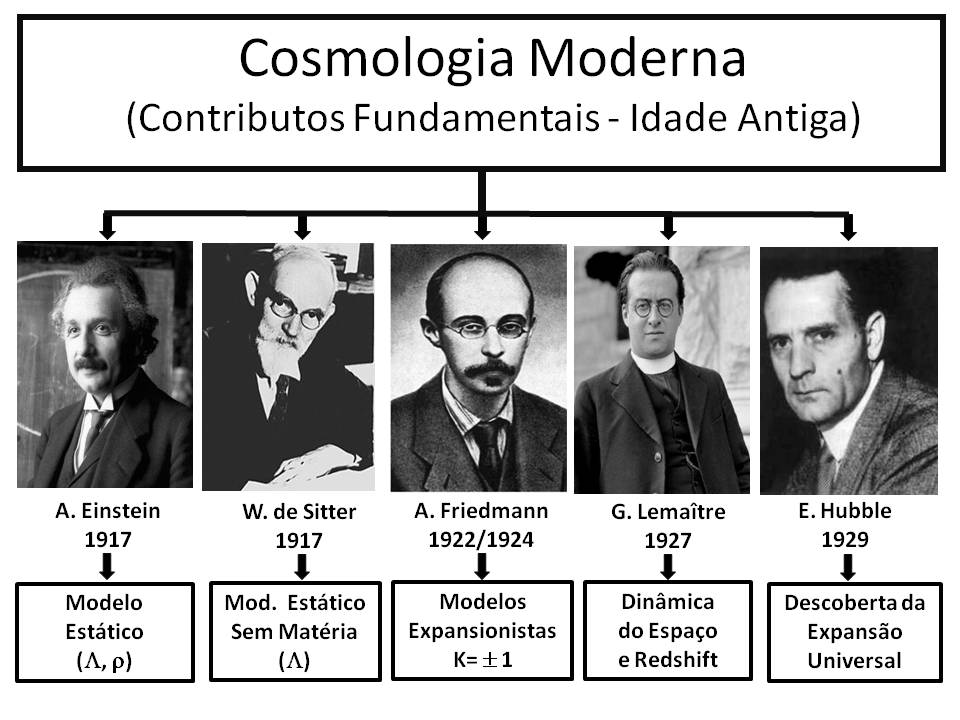}
\caption{Os fundadores da cosmologia. O quarteto fantástico (Einstein, de Sitter, Friedmann e Lemaître)  resolveu os principais desafios teóricos da época, enquanto Hubble fundou a astronomia extragaláctica e descobriu a expansão do Universo.}\label{Contributos}
\end{figure}

\section{Conclusão da \textit{Parte I}}

Durante mais de 2 séculos, nada de substancial foi acrescentado a vis\~ao teórica newtoniana de um universo est\'atico, infinito e euclidiano,  regido por sua auto-gravita\c{c}\~ao (cf. seção 2).   

De posse deste legado, Einstein demorou de 1907 à 1915 para obter a relatividade geral (uma descrição da gravitação mais precisa do que a teoria newtoniana) e apenas dois anos para iniciar a cosmologia relativística com seu modelo estático de 1917 (ver \textbf{Figura 3}). No mesmo ano,  de Sitter obteve  uma segunda solução estática (sem matéria) e apenas 5 anos depois (1922), Friedmann obteve as primeiras soluções expansionistas para o caso fechado com e sem $\Lambda$, repetindo para o caso aberto (k=-1) em 1924. Como visto na seção anterior, as importantes contribuições de Lemaître (1927),  repercutiram apenas depois 1931. Mesmo assim,  Edwin Hubble, o pai da astronomia extragaláctica descobriu  a expansão do Universo em 1929.

Na \textbf{Figura 10}, mostramos os pioneiros da cosmologia moderna e suas maiores contribuições até a descoberta da expansão. 

É importante enfatizar que Hubble estava consciente da possível conexão entre sua descoberta e as idéias relativísticas, embora não tenha citado os trabalhos de Friedmann e de Sitter explicitamente (ver seção 7 e nota de rodapé 37). No entanto, foi Lemaître, em 1927, quem provou que o desvio para o vermelho das linhas espectrais é uma consequência natural da expansão do Universo. Um trabalho bastante divulgado a partir da tradução publicada na MNRAS, em 1931 \cite{L31}.

Seguindo Lemaître, os cosmólogos  sempre mantiveram  a mente aberta, considerando a existência de $\Lambda$. Essa constante universal ainda desconhecida, poderia  resolver o problema da idade. Foi a primeira ressurreição de $\Lambda$  dentre muitas em 100 anos de cosmologia moderna. A mais recente será vista na parte III, tendo sido causada exatamente pela descoberta da aceleração do Universo. 

\textbf{Agradecimentos:} JASL agradece ao CNPq,  CAPES (PROCAD 2013) e   FAPESP (Projeto LLAMA).    

\vskip 0.5cm
\noindent \textbf{\large Apêndice A: Fator de Escala,  \textit{Redshift} e a Lei de Hubble}
\vskip 0.5cm
Em cosmologia todas as  quantidades que dependem do tempo, podem ser expressas como funções do \textit{redshift}. Neste apêndice deduziremos  uma  expressão para o fator de escala $a(t)$ como função dependente do \textit{redshift} $z$ \cite{W2008,L31}.  

Fótons vindos de uma fonte distante seguem geodésica nulas ($ds^{2}=0$). Por simetria, podemos escolher na métrica de FRW \eqref{FRW},  $d\theta=d\phi=0$, obtendo:  
\begin{equation}\label{A1}
\nonumber
 ds^{2}=0 \,\,\,\,\Longleftrightarrow\,\,\,\,\,  cdt = -\frac{a(t)dr}{1-{kr^{2}}} .\,\,\,\,\,\,\,\,\,\,\,\,\,\,\,\,\,\,\,\,\,\,\,\,\,\,\,\,\,\,\,\,\,(A1)
\end{equation}

Considere uma frente de onda monocromática emitida pela fonte  na posição $r$ no instante  $t$ e chegando na origem $r=0$ em $t=t_0$. A integral da equação acima é dada por: 

\begin{equation}\label{A2}
\nonumber
\int^{t_o}_{t} \frac{cdt}{a(t)} = \int^{r}_{o} \frac{dr}{1-kr^{2}}=F(r)\,. \,\,\,\,\,\,\,\,\,\,\,\,\,\,\,\,\,\,\,\,\,\,\,\,\,\,\,\,\,\,\,\,\,\,\,\,\,\,\,\,\,\,(A2)
\end{equation}
A segunda crista, emitida em $t+\Delta t$, chegará no detector em $t_0 + \Delta t_0$ e como a coordenada $r$ (comóvel) da fonte é fixa podemos escrever:
\begin{equation}\label{A3}
\nonumber
\int^{t_o + \Delta t_o}_{t + \Delta t} \frac{cdt}{a(t)} = F(r) = \int^{t_o}_{t} \frac{cdt}{a(t)} . \,\,\,\,\,\,\,\,\,\,\,\,\,\,\,\,\,\,\,\,\,\,\,\,\,\,\,\,\,\,\,\,\,\,\,\,\,\,\,\,\,(A3)
\end{equation}
Considerando que  os períodos emitidos na fonte e recebidos pelo observador na origem são, respectivamente, $\Delta t$ e $\Delta t_0$  e  definindo ${\Delta \lambda}=\lambda_0 - \lambda$, temos:  
\begin{equation}\label{A4}
\nonumber
\frac{\Delta t_0}{\Delta t} = \frac{\lambda_0}{\lambda}=\frac{a_0}{a} \Longleftrightarrow \frac{\Delta \lambda}{\lambda}\equiv z=\frac{a_0}{a} -1\,,   \,\,\,\,\,\,\,\,\,\,\,\,\,\,\,\,\,\, \,\,\,(A4)
\end{equation}
ou ainda:
\begin{equation}\label{A5}
\nonumber
\hskip 0.8cm a(z) = \frac{a_0}{1+z}\,\,.\,\,\,\,\,\, \,\,\,\,\,\,\,\,\,\,\,\,\,\,\,\,\,\,\,\,\,\,\,\,\,\,\,\,\,\,\,\,\,\,\,\,\,\,\,\,\,\,\,\,\,\,\,\,\,\,\,\,\,\,\,\,\,\,\,\,\,\,\,\,\,\,\,\,\,\,\,\,\,\,\,\,\,\,\,\,\,\,\,\,\,\,\,\,\,\,\,\,\,\,\,\,\,(A5)
\end{equation}
Note que hoje ($z=0$) temos $a=a_0$ e quando $z\rightarrow \infty$, $a\rightarrow 0$ (singularidade).  Como o volume comóvel, $V\propto a^{3}$, temos  que em $z=1$, $V(z=1)/V_0$ = 1/8. A expressão acima  é muito útil em cosmologia, pois todas as quantidades que dependem do fator de escala (ou do tempo), tais como densidade, temperatura, etc.,  podem ser imediatamente escritas em termos de $z$\,. Em particular, a densidade de matéria não-relativística, $\rho \propto a^{-3}$, é usualmente expressa como: $\rho = \rho_0(1+z)^{3}$, onde $\rho_0$ é a densidade em $z=0$. Como esperado, $\rho (z)$ diverge na singularidade ($z \rightarrow \infty$).

A expressão  (A4) acima foi primeiramente deduzida por Lemaître em 1927. A partir dela é imediato obter a lei de Hubble para pequenas distâncias (baixos redshifts), outro resultado de Lemaître. Temos:

\begin{equation}
\nonumber
z=\frac{V}{c} = \frac{a_0 - a}{a}= \frac{da}{a} = \frac{\dot a}{a}dt=Hdt\,, \,\,\,\,\,\,\,\,\,\,\,\,\,\,\,\,\,\,\,\,\,\,\,\,\,\,\,\,\,\,\,\,\,\,(A6) 
\end{equation}
ou ainda,
\begin{equation}
\nonumber
V =H\,cdt\simeq H(t)r. \,\,\,\,\,\,\,\, \,\,\,\,\,\,\,\,\,\,\,\,\,\,\,\,\,\,\,\,\,\,\,\,\,\,\,\,\,\,\,\,\,\,\,\,\,\,\,\,\,\,\,\,\,\,\,\,\,\,\,\,\,\,\,\,\,\,\,\,\,\,\,\,\,\,\,\,\,\,\,\,\,\,\,\,\,\,\,\,\,\,\,\,\,\,\,\,\,\,\,\,\,(A7)
\end{equation}
Finalmente, expandindo $H(t)$ em série de Taylor em torno de $t=t_0$ e retendo apenas o primeiro termo, segue a  lei de Hubble, $V=H_0\,r$.

\end{document}